\documentclass[a4paper,11pt]{article}
\usepackage{amssymb,amsmath,amsthm}
\usepackage{a4wide}
\usepackage{pgf,tikz}
\usetikzlibrary{arrows}
\usepackage{epsfig}
\usepackage{color}
\usepackage{mathdots}

\usepackage{hyperref}

\usepackage{authblk}

\usepackage{subcaption}

 \newcommand{\sfrac}{\genfrac{}{}{}2}

\newcommand{\beq}{\begin{equation}}  
\newcommand{\eeq}{\end{equation}}  
\newcommand{\bea}{\begin{eqnarray}} 
\newcommand{\eea}{\end{eqnarray}}   
\newcommand{\bear}{\begin{array}}  
\newcommand{\eear}{\end{array}}

\newtheorem{thm}{Theorem}[section] 

\newtheorem{propn}[thm]{Proposition}

\newtheorem{lem}[thm]{Lemma}

\newtheorem{cor}[thm]{Corollary} 

\newenvironment{prf}{\trivlist \item [\hskip 
\labelsep {\bf Proof:}]\ignorespaces}{\qed \endtrivlist}

\theoremstyle{definition}

\newtheorem{exa}[thm]{Example}
\newtheorem{remark}[thm]{Remark}

\newcommand{\Z}{{\mathbb Z}}
\newcommand{\C}{{\mathbb C}}
\newcommand{\R}{{\mathbb R}}

\newcommand{\rd}{\mathrm{d}}
\newcommand{\ri}{\mathrm{i}}

\newcommand\La{{\Lambda}}

\newcommand\al{{\alpha}}
\newcommand\be{{\beta}}
\newcommand\ze{{\zeta}}
\newcommand\gam{{\gamma}}

\newcommand\om{{\omega}}

\newcommand\si{{\sigma}}

\begin{document}


\title{Similarity reductions of peakon equations: integrable cubic equations}
\author[1]{L.E. Barnes} 
\author[1]{A.N.W. Hone} 
\author[2]{M. Senthilvelan}
\author[2]{S. Stalin} %
\affil[1]{School of Mathematics, Statistics \&  Actuarial Science, 
University of Kent, 
Canterbury CT2 7FS, U.K. 
e-mail: A.N.W.Hone@kent.ac.uk
}
\affil[2]{Department of Nonlinear Dynamics, Bharathidasan University, 
Tiruchirappalli  620 024, Tamilnadu, India.
}

\maketitle

\begin{abstract} 
We consider 
the scaling similarity solutions of two integrable cubically nonlinear partial differential equations (PDEs)
that admit peaked soliton (peakon) solutions, namely the modified Camassa-Holm (mCH) equation and Novikov's equation. 
By making use of suitable reciprocal transformations, which map the mCH equation and Novikov's equation 
to a negative mKdV flow and a negative Sawada-Kotera flow, respectively, we show 
that each of these scaling similarity reductions is related via a hodograph transformation to an equation of Painlev\'e type: 
for the mCH equation, its reduction is of second order and second degree, while for Novikov's equation the reduction is a particular case 
of 
Painlev\'e V. 
Furthermore, we show that each of these two different Painlev\'e-type equations is  
related to the particular cases of Painlev\'e III that arise from analogous 
similarity reductions of   
the Camassa-Holm and the Degasperis-Procesi equation, respectively. 
For each of the cubically nonlinear PDEs considered, we also give explicit parametric forms 
of their periodic travelling wave solutions in terms of elliptic functions. We present some parametric plots of the latter, 
and, by using explicit algebraic solutions of Painlev\'e III, we do the same for some of the simplest  examples of scaling similarity solutions, together with descriptions of their leading order 
asymptotic behaviour.
\end{abstract}

\section{Introduction}

\setcounter{equation}{0}

\subsection{Background and motivation} 

Painlev\'e transcendents  can naturally be regarded as nonlinear analogues of the classical special functions. Classical special functions, such as Legendre polynomials, Hermite polynomials, or Bessel functions, which satisfy linear ordinary differential equations (ODEs), arise in the solution of linear 
partial differential equations (PDEs) by the method of separation of variables. In a similar way, Painlev\'e transcendents, which satisfy 
nonlinear ODEs, provide similarity solutions of soliton-bearing PDEs that are solvable by the inverse scattering transform \cite{ars, fn}, and are now known 
to describe universal features of critical behaviour in such nonlinear PDEs (see e.g.\ \cite{dgk}), as well 
as appearing in the treatment 
of scaling phenomena and other aspects of random matrices, statistical mechanics and 
quantum field theories (see \cite{h} and chapter 32 in \cite{dlmf} for references to these and various other applications). 
The aim of this paper is to explain how, in a somewhat indirect way,  Painlev\'e equations appear in the analysis of 
scaling similarity reductions of certain integrable nonlinear PDEs that admit peaked soliton solutions (peakons). 

The cubically nonlinear 
PDE given by 
\beq\label{mch} 
u_t-u_{xxt} +3u^2u_x+2u_xu_{xx}^2+u_x^2u_{xxx}=u_x^3+4uu_xu_{xx}+u^2u_{xxx}
\eeq 
is commonly known as the modified Camassa-Holm (mCH) equation (see e.g.\ \cite{gui}, or \cite{cs, wlm}), because it is related via a reciprocal transformation to a negative flow in the modified KdV hierarchy, 
while the Camassa-Holm equation \cite{ch1}, that is 
\beq\label{ch}
u_t-u_{xxt} +3uu_x=2u_xu_{xx}+uu_{xxx},
\eeq 
has an analogous relationship with 
a corresponding negative flow in the KdV hierarchy. 
These and similar equations arise as truncations of asymptotic series approximations in shallow water theory \cite{ch2, cl, dgh1, dgh2, ivanov}, as 
bi-Hamiltonian equations admitting infinitely many commuting symmetries generated by a recursion operator \cite{ff, mn, or}, and as 
compatibility conditions coming from a Lax pair \cite{fokas, qiao}. In these various contexts, the equation (\ref{ch}) can appear 
with additional linear dispersion ($u_x$ and $u_{xxx}$) terms, while (\ref{mch}) can have suitable linear and quadratic nonlinear terms included, 
but such terms can always be removed by a combination of a Galilean transformation and a shift $u\to u+\,$const, which changes the boundary 
conditions. However, such transformations are irrelevant from the point of view of integrability, 
which is determined by the underlying algebraic structure of the equations and their symmetries, 
so here we will always work with the pure dispersionless versions of 
these equations. 

Aside from their connections with shallow water models and bi-Hamiltonian theory, 
perhaps the most remarkable feature of the dispersionless forms of these equations, 
as discovered  in \cite{ch1} for  (\ref{ch}),
is 
the fact that they admit weak solutions in the form of peaked solitons with discontinuous derivatives at the peaks, given by 
\beq\label{peakons} 
u(x,t)=\sum_{j=1}^N p_j(t) \exp\big(-|x-q_j(t)|\big), 
\eeq 
where  $q_j(t)$ are the positions of the peaks and $p_j(t)$ are the amplitudes, which  satisfy a finite-dimensional Hamiltonian system of ODEs. 
Due to this feature, we refer to these PDEs as peakon equations. The 
characteristic shape of 
the peakons can be understood by introducing   the momentum density $m$, given by 
the 1D Helmholtz operator acting on the velocity field $u$, that is 
\beq\label{mom}
m=(1-D_x^2)u.
\eeq 
The introduction of $m$ allows the dispersionless versions of the PDEs to be rewritten in a very compact form: 
the Camassa-Holm equation (\ref{ch}) is equivalent to
\beq\label{chm}
m_t+um_x+2u_xm=0, 
\eeq 
while the modified equation (\ref{mch}) is rewritten in the simple form 
\beq\label{mchm}
m_t+\big(m(u^2-u_x^2)\big)_x=0.
\eeq  
Then since $\tfrac{1}{2}e^{-|x|}$ is the Green's function for the Helmholtz operator, 
the momentum density (\ref{mom}) for the peakon solutions (\ref{peakons}) is a sum of Dirac 
delta functions with support at the peak positions $q_j(t)$, $j=1,\ldots,N$. 
For more details  of peakons and their connections with approximation theory, 
see the recent review \cite{ls3} and references therein. 

The other cubically nonlinear PDE we will be concerned with here is the equation 
\beq\label{novikov}
u_t-u_{xxt}+4u^2u_x=u^2u_{xxx}+3uu_xu_{xx}, 
\eeq 
which was found by Novikov in a classification of quadratic/cubic peakon-type equations  admitting infinitely 
many symmetries \cite{novikov}. Novikov's equation can be written more concisely as 
\beq\label{vn} 
m_t+u^2m_x+3uu_xm=0, 
\eeq 
with the same momentum variable as in  (\ref{mom}). As shown in \cite{hw2}, it is related by a reciprocal transformation 
to a negative flow in the Sawada-Kotera hierarchy; so its relationship with the 
Degasperis-Procesi equation  \cite{dp}
\beq\label{dp} 
m_t+um_x+3u_xm=0, 
\eeq 
which has a reciprocal transformation to a negative Kaup-Kupershmidt flow \cite{dhh1}, is somewhat similar to the 
relationship between (\ref{mch}) and (\ref{ch}), because the Sawada-Kotera and Kaup-Kupershmidt hierarchies 
are related to the same modified hierarchy (see \cite{fg, pa} and references).

In this paper we are concerned with scaling similarity solutions of the cubic peakon equations (\ref{mch}) and 
(\ref{novikov}). It is a surprising feature of all the peakon equations described so far that, despite being fundamentally nonlinear, they admit 
separable solutions of the form 
$$ 
u(x,t)=Y(t)U(x), 
$$ 
which are typically only a feature of linear PDEs. It turns out that (up to the trivial freedom to shift $t$ by a constant, which 
henceforth we will ignore)  the function $Y$ is just a power of $t$, corresponding to a 
simple symmetry of each of these equations under scaling $u$ and $t$. 
For the cubic peakon equations  being considered here, the separable solutions take the form 
\beq\label{sep} 
u(x,t) =t^{-1/2} U(x), 
\eeq 
while for the quadratically nonlinear equations (\ref{ch}) and (\ref{dp}) these solutions have the form 
\beq\label{sepch} 
u(x,t) =t^{-1} U(x)
\eeq 
instead. 

Solutions of the form (\ref{sepch}) for the 
Camassa-Holm equation (\ref{ch}) were discussed in \cite{gp}, 
where it was shown that the ODE satisfied by $U(x)$ fails the Painlev\'e test. From this point of view, 
it would appear that such solutions are a counterexample to an 
assertion of Ablowitz, Ramani and Segur \cite{ars}, which says that all ODEs obtained from 
similarity reductions of PDEs that are integrable (in the sense of admitting a Lax pair, 
so that they can be solved by the inverse scattering method) should be free of movable critical 
points. However, it has long been known that this assertion cannot be true in its most naive form, 
and the Camassa-Holm equation is a case in point: the 
expansion of the solutions of the PDE (\ref{ch}) itself in the neighbourhood of a movable singularity manifold 
displays algebraic branching \cite{h}, 
and this movable branching is inherited by the ODEs obtained from it via a similarity reduction, such as the 
equation for $U(x)$. Nevertheless, the solutions (\ref{sepch}) belong to a one-parameter family of 
similarity reductions, found in \cite{hach}, which can be solved in terms of certain 
Painlev\'e III transcendents, meaning that the assertion of \cite{ars} can be salvaged in this case. 
Although this would appear to contradict the result of \cite{gp}, there is in fact no contradiction: 
the Painlev\'e property, and more specifically the third Painlev\'e equation 
\beq\label{piii}
\frac{\rd^2 w}{\rd \ze^2}=\frac{1}{w}\left(\frac{\rd w}{\rd \ze}\right)^2-\frac{1}{\ze}\left(\frac{\rd w}{\rd \ze}\right)+\frac{1}{\ze}\big( \tilde{\al} w^2+\tilde{\be}\big) +\tilde{\gam} w^3 +\frac{\tilde{\delta}}{w} 
\eeq
(for some particular values of the parameters  $\tilde{\al} ,  \tilde{\be},  \tilde{\gam},  \tilde{\delta}$), 
only appears after making certain 
precise changes of the dependent and independent variables, 
including a hodograph-type transformation, which completely changes the  singularity structure. Thus 
movable poles of $w$  in the complex $\ze$ plane arise from movable algebraic branch points  in terms of the     
original variables, i.e.\ $U$ and $x$ in (\ref{sepch}).

In recent work \cite{bh}, we studied similarity reductions of the so-called $b$-family of equations, given by 
\beq\label{bfam} 
m_t +um_x+bu_xm=0, \qquad m=u-u_{xx}, 
\eeq 
with a constant coefficient $b$.
It is known that the cases $b=2,3$, namely (\ref{ch}) and (\ref{dp}), are the only members of this family that 
are integrable in the sense of admitting infinitely many commuting local symmetries \cite{mn}, and the only cases 
for which the prolongation algebra method provides a Lax pair of zero curvature type \cite{hw1}.  For any $b$, the equation 
(\ref{bfam}) admits  
a one-parameter family of scaling similarity reductions which includes the separable solutions (\ref{sepch}), and in particular
the ramp profile 
\beq\label{ramp} 
u(x,t)=\frac{x}{(b+1)t} 
\eeq 
for $b\neq -1$.  Beginning with \cite{hs1},
Holm and Staley did extensive studies on  numerical solutions of (\ref{bfam}),  and revealed 
bifurcation phenomena controlled by the parameter $b$. Their results  included numerically stable 
``ramp-cliff'' solutions for $-1<b<1$, 
looking like the 
ramp (\ref{ramp}) 
in a compact region, joined to a rapidly decaying cliff. The results in \cite{bh} show that the scaling similarity reductions 
of (\ref{bfam}) are  related via a  transformation of hodograph type to a 
non-autonomous ODE of second order; but this ODE only has 
the Painlev\'e property in the integrable cases $b=2,3$, when it is  equivalent to two different versions of the 
Painlev\'e III  equation (\ref{piii}): the reduction already found for the Camassa-Holm equation in \cite{hach}, 
and another set of values of $\tilde{\al} ,  \tilde{\be},  \tilde{\gam},  \tilde{\delta}$
for the reduction of the Degasperis-Procesi equation. 

\subsection{Outline of the paper}
An outline of the rest of the paper is as follows. 

The next section is devoted to similarity reductions of the 
mCH equation (\ref{mch}). We begin by briefly reviewing the link between the Camassa-Holm equation 
and the first negative KdV flow via a  reciprocal transformation, as well as the reciprocal 
transformation between (\ref{mch}) and the first negative mKdV flow, before presenting   
a precise formulation of the Miura map between these two negative flows (Proposition \ref{miura}). 
Our main goal is then to describe the scaling similarity solutions of (\ref{mch}), 
but to pave the way towards this it is helpful to consider the travelling wave solutions  
beforehand. The latter are related to corresponding travelling waves of 
the first negative mKdV flow via a  transformation of  hodograph type, 
which is obtained by applying the travelling wave reduction to the reciprocal 
transformation. By reduction of the associated Miura map, these solutions 
are then connected to explicit elliptic function formulae for the travelling waves of the  
negative KdV equation, as found in \cite{hach}. This leads 
to an exact parametric solution for the smooth  travelling waves of (\ref{mch}), in terms 
of Weierstrass functions, given in Theorem 
\ref{mchtrav} below, and illustrated with plots of a particular solution (see Example \ref{mchtravplot}). 
The same template is followed for the scaling similarity solutions of (\ref{mch}): the similarity reduction of  
the reciprocal transformation provides a hodograph-type link between these solutions and a Painlev\'e-type ODE of 
second order and second degree for corresponding solutions of the first negative mKdV equation; 
and the reduction of the Miura map gives  a one-to-one correspondence 
between the second degree equation and the particular case of Painlev\'e III 
that is associated with the scaling similarity solutions of  the negative KdV flow (Lemma \ref{corr}). 
The main result of the section is the parametric form of the scaling similarity solutions of (\ref{mch}), given 
in terms of a solution of the second degree equation and a pair of tau functions for Painlev\'e III (Theorem \ref{mchmain}). 
An explicit  illustration of this result, together with the leading order 
asymptotics of two real branches in a particular similarity solution, is provided in Example \ref{Uzex}, which is based 
on a simple algebraic solution of  Painlev\'e III.

Section 3 is concerned with similarity reductions of Novikov's equation (\ref{novikov}).  
Initially, we review the two different Miura maps that relate the Kaup-Kupershmidt hierarchy and 
the Sawada-Kotera hierarchy to the same modified hierarchy, as well as the 
negative flows in each of these hierarchies which are linked via a reciprocal transformation 
to the Degasperis-Procesi equation (\ref{dp}), and to Novikov's equation, respectively. 
Once again, to lay the groundwork for the subsequent results on scaling similarity 
solutions, it is helpful to first make a detailed analysis of the travelling waves for 
(\ref{novikov}). By reduction of the reciprocal transformation connecting it to 
the negative Sawada-Kotera equation, these are related to the travelling wave 
solutions of the latter, which are given 
explicitly in terms of elliptic functions; hence the exact parametric form of 
the smooth travelling waves in Novikov's equation is derived (Theorem \ref{vntrav}), 
and a particular numerical example is plotted (Example \ref{vntravplot}). 
The analysis of the scaling similarity solutions follows a similar pattern: in this case, the reciprocal transformation 
reduces to a  hodograph link with the solutions of a non-autonomous ODE of second order, which describes the 
corresponding scaling similarity reduction of the negative Sawada-Kotera flow. After a simple change 
of variables, this ODE is shown to be equivalent to a case of 
the fifth Painlev\'e equation, that is
\beq\label{pV}
\frac{\rd^2 w}{\rd \ze^2}=\left(\frac{1}{2w}+\frac{1}{w-1}\right)\left(\frac{\rd w}{\rd \ze}\right)^2-\frac{1}{\ze}\left(\frac{\rd w}{\rd \ze}\right)+\frac{(w-1)^2}{\ze^2}\Big( \tilde{\al} w+\frac{\tilde{\be}}{w}\Big) 
+\frac{\tilde{\gam} w}{\ze} +\frac{\tilde{\delta} w(w+1)}{w-1},
\eeq
with a particular restriction on the parameters, including the requirement that 
$\tilde{\delta}=0$. By a result due to Gromak 
\cite{gromak}, 
when this requirement holds, the Painlev\'e V equation (\ref{pV})  
is solved with Painlev\'e III transcendents, and we use this 
to give a one-to-one correspondence between the ODEs for the scaling similarity solutions 
of the negative Sawada-Kotera and Kaup-Kupershmidt flows
(Proposition \ref{PRprop}).   
These ODEs have B\"acklund transformations, which 
can be deduced from the action of certain discrete symmetries that are inherited from the Miura maps 
for the two PDE hierarchies  (see Corollary \ref{PodeBTs}). The general scaling similarity solution 
of Novikov's equation is then given parametrically in terms  of a solution 
of the aforementioned ODE that is equivalent to a case of Painlev\'e V, together with two different 
tau functions related by a B\"acklund transformation (Theorem \ref{vnscalparam}). 
As in the case of the reductions of the mCH equation, an illustration of the latter result is provided 
by starting from an elementary algebraic solution of Painlev\'e III (Example \ref{vnsimexa}), 
for which we plot the corresponding  scaling similarity solution of (\ref{novikov}), and determine 
its leading order asymptotics for large positive/negative real values of the independent 
variable. 
To conclude the section, we consider the special case of the 
separable solutions of the form (\ref{sep}) in Novikov's equation, 
which turn out to be given parametrically in terms of two different 
quadratures involving Bessel functions of order zero 
(Theorem \ref{alzero}). 

The fourth and final section of the paper contains our conclusions. 

Preliminary versions of some of these results on scaling similarity reductions of cubic peakon equations were 
presented in the thesis \cite{lucy}. 

\section{Reductions of the mCH 
equation}

\setcounter{equation}{0}

In this section, we consider reductions of the modified Camassa-Holm (mCH) equation (\ref{mch}). To begin 
with, we explain why the latter nomenclature is appropriate, by describing the relationship 
with the Camassa-Holm equation (\ref{ch}), which becomes apparent when suitable reciprocal transformations 
are applied to these two equations. In order to clearly distinguish between the two equations, we start from the 
Camassa-Holm equation in the form (\ref{chm}), and write the associated 
conservation law for the field $p=\sqrt{m}$, while at the same 
time replacing the other dependent/independent variables $u,x,t$ by $\bar{u},\bar{x},\bar{t}$, so that it becomes the 
system 
\beq\label{chbar} 
\frac{\partial p}{\partial \bar{t}}+\frac{\partial }{\partial \bar{x}}(\bar{u}p)=0, \qquad 
p^2=(1-D_{\bar{x}}^2)\bar{u}.
\eeq 
Henceforth in this section we reserve $m,u,x,t$ for the corresponding dependent/independent variables in the 
modified equation (\ref{mchm}).

\subsection{Miura map between negative flows}

The first equation in (\ref{chbar}) is a conservation law for the Camassa-Holm equation, which leads to the 
introduction of new independent variables $X,T$ via the reciprocal transformation 
\beq\label{chrt}
\rd X=p\,\rd \bar{x} -\bar{u}p\,\rd \bar{t},\qquad \rd T=\rd \bar{t}.
\eeq 
By applying a reciprocal transformation to a PDE system, any conservation law in the original independent variables 
is transformed to another conservation law in terms of the new variables. For the Camassa-Holm equation, the result of 
applying the reciprocal transformation (\ref{chrt}) is a PDE of third order for $p=p(X,T)$, which can be written in 
conservation form as
\beq\label{p3rd}
\frac{\partial }{\partial T} (p^{-1})  +  \frac{\partial }{\partial X}\big(p (\log p)_{XT}-p^2\big)=0.
\eeq  
(Here and throughout the rest of the paper, we abuse notation by using the same letter to denote a field variable 
as a function of both old and new independent variables, so $p(x,t)\to p(X,T)$.)  
An alternative way to express the equation (\ref{p3rd}) in conservation form, which makes the connection 
with the KdV hierarchy apparent, is 
\beq\label{VT}
\frac{\partial V}{\partial T}+\frac{\partial p}{\partial X}=0, 
\eeq    
where the quantity $V$ is defined in terms of $p$ by 
\beq\label{ep}
pp_{XX}-\tfrac{1}{2}p_X^2 +2Vp^2+\tfrac{1}{2}=0.
\eeq 
The quantity $V$ is the usual KdV field variable, which (up to scale) appears in the Lax pair as the potential in a 
Schr\"odinger operator, and it follows from (\ref{VT}) and (\ref{ep}) that ${\cal R}V_T=0$, where 
${\cal R}=D_X^2 +4V +2V_XD_X^{-1}$ is the recursion operator for the KdV hierarchy. 
Hence the PDE (\ref{p3rd}) obtained by applying the above reciprocal transformation to the Camassa-Holm equation 
corresponds to the first negative KdV flow (see  \cite{hw1} and references therein for further discussion).

An analogous reciprocal transformation for the modified equation (\ref{mchm}) is defined 
by 
\beq\label{mchrt}
\rd X =\tfrac{1}{2}m\,\rd x-\tfrac{1}{2}fm\,\rd t, \qquad \rd T =4\,\rd t,
\eeq   
with 
\beq\label{fdef}
f=u^2-u_x^2.
\eeq 
The latter transformation (with $T$ rescaled) was presented in \cite{hw2}, where the connection with the 
modified KdV (mKdV) hierarchy  was obtained by deriving the standard Miura map formula 
from the reciprocal transformation applied to the mCH Lax pair in the form given by Qiao \cite{qiao}. 
Here we make this connection more explicit, and we shall see 
that the choice of scale factor 4 in the definition of $T$ is important in what follows.

Direct application of the  transformation (\ref{mchrt}) to the modified Camassa-Holm equation  (\ref{mchm}) 
results in the conservation law 
\beq\label{vT}
v_T=\tfrac{1}{8}f_X, 
\eeq
where it is convenient to introduce the field 
\beq\label{vdef}
v=m^{-1}.
\eeq 
In order to obtain a single PDE for $v=v(X,T)$, it is necessary to make use of the definitions (\ref{fdef}) 
and (\ref{mom}). These 
yield 
\beq\label{fx}
f_x=2u_xm \implies vf_X=2u_X 
\eeq
and 
$$ 
m=u-u_{xx}\implies
v^{-1}=u-\tfrac{1}{4}\big(v^{-1}D_X\big)^2 u 
=u-\tfrac{1}{8}v^{-1}D_X(f_X)=u-v^{-1}v_{XT},
$$ 
where (\ref{vT}) was used to obtain the last equality, which rearranges to produce 
\beq\label{ueq}
u=v^{-1}(v_{XT}+1).
\eeq
Then from (\ref{fx}) and (\ref{vT}) we have another conservation law, that is 
\beq\label{vsqT}
\frac{\partial }{\partial T}(v^2)=\tfrac{1}{2}u_X,
\eeq 
and by substituting for $u$ from (\ref{ueq}) in the right-hand side above, this gives a PDE of third order for $v$, 
namely 
\beq\label{veqn}
\frac{\partial }{\partial X}\Big(v^{-1}(v_{XT}+1)\Big) =4vv_T.
\eeq 

The equation (\ref{veqn}) was not explicitly given in \cite{hw2}, where the result of the reciprocal transformation applied to 
(\ref{mchm}) was instead written as a system, while the interpretation of this as a negative mKdV flow was inferred 
from the transformation of the Lax pair, revealing that the KdV field $V$ in (\ref{VT}) is given in terms of $v$ by the 
standard Miura relation 
\beq\label{Vv}
V=v_X-v^2.
\eeq  
We now describe the relation between (\ref{p3rd}) and (\ref{veqn}) more precisely. 
 
\begin{propn}\label{miura}
The first negative mKdV flow, given by (\ref{veqn}), is mapped to the first negative KdV flow (\ref{p3rd}) by the Miura 
transformation 
\beq\label{pv}
p=\tfrac{1}{2}v^{-1}(v_{XT}+1)-v_T,
\eeq 
which has the Miura map (\ref{Vv}) for the KdV field $V$ as a consequence. 
Conversely, if $p=p(X,T)$ is a solution of the PDE (\ref{p3rd}), then $v=v(X,T)$ defined by 
\beq\label{vpinv} 
v=-\tfrac{1}{2}p^{-1}(p_X-1),
\eeq 
satisfies the PDE  (\ref{veqn}). 
\end{propn}
\begin{prf}
Using (\ref{ueq}), the formula (\ref{pv}) can be rewritten as 
$$ 
p=\tfrac{1}{2}u-v_T,
$$ 
so if $v=v(X,T)$ is a solution of the equation (\ref{veqn}) then 
\beq\label{pX} 
p_X=\tfrac{1}{2}u_X-v_{XT}=2vv_T-v_{XT}, 
\eeq 
by (\ref{vsqT}). Then 
upon rearranging (\ref{pv}), we find 
\beq\label{vpXT}
2vp-1=v_{XT}-2vv_T, 
\eeq 
which means that applying the identity (\ref{pX}) 
yields 
$$ 2vp-1
=-p_X, 
$$
and hence $v$ can be written in terms of $p$, in the form (\ref{vpinv}). 
This expression for $v$ 
then gives 
\beq\label{Vcalc}
\begin{array}{rcl}
v_X-v^2&=&\tfrac{\partial}{\partial X} \big(-\sfrac{1}{2}p^{-1}(p_X-1)\big) -\sfrac{1}{4}p^{-2}(p_X-1)^2 \\ 
& = & -\sfrac{1}{2}p^{-1}p_{XX}+\sfrac{1}{4}p^{-2}(p_{X}^2-1), 
\end{array}  
\eeq  
where the expression on the last line is the definition of the KdV field $V$ in terms of $p$, according to (\ref{ep}); 
thus 
$V$ is given in terms of $v$ by the standard Miura formula (\ref{Vv}). Differentiating the latter formula 
with respect to $T$ produces 
\beq\label{MiuraT}
V_T=v_{XT}-2vv_T,  
\eeq 
which implies $V_T=-p_X$,
by using  (\ref{pX}) once more. Thus $p=p(X,T)$ defined by (\ref{pv}) satisfies (\ref{VT}), which is equivalent to 
the PDE (\ref{p3rd}). 
For the converse, if $v$ is given in terms of a solution of (\ref{p3rd}) by (\ref{vpinv}), 
then the calculation (\ref{Vcalc}) giving the Miura formula for $V$ in terms of $v$ holds, and its $T$
derivative yields the equality (\ref{MiuraT}). 
Hence, by applying (\ref{p3rd}), $v_{XT}-2vv_T=-p_X$, or equivalently 
\beq\label{pXvXT} 
p_X+v_{XT}=2vv_T, 
\eeq 
and then from (\ref{vpinv}) this implies 
that (\ref{vpXT}) holds, which in turn means that $p$ can be written in terms of $v$ according to (\ref{pv}). 
Finally, differentiating both sides of (\ref{pv}) with respect to $X$ and using this to substitute for $p_X$ in 
(\ref{pXvXT}), the PDE (\ref{veqn}) for $v$ follows. 
\end{prf} 
The preceding result shows exactly why 
``modified Camassa-Holm equation'' is a suitable name for (\ref{mchm}), since under a 
reciprocal transformation it is connected to (\ref{chm}) by a Miura map.

\subsection{Travelling waves} 

Before treating the scaling similarity solutions, 
we first consider travelling waves of the mCH equation (\ref{mchm}), setting 
\beq\label{travwaves}
u(x,t)=U(z), \qquad m(x,t)=M(z), \qquad z=x-ct, 
\eeq 
where $c$ is the wave velocity, and we will also write $F(z)$ for the quantity $U^2-U_z^2$ obtained 
from (\ref{fdef}). 
As it is already in the form of a conservation 
law, (\ref{mchm}) becomes a total $z$ derivative, so integrating this we obtain 
\beq\label{up}
(F-c)M+k=0,  
\eeq
where $k$ is an integration constant. 
Henceforth we will assume that $k\neq 0$, since 
if we are considering 
smooth solutions,  
then the case $k=0$ implies that either $F=c$, or $M=0$, 
both of which lead to unbounded solutions given in terms of exponential/hyperbolic functions; 
but the 1-peakon solution with $M$ being given by a delta function can 
be viewed as a weak limit of strong (analytic) solutions with $k=0$ \cite{lo}. 
In terms of $z$ derivatives, the first equality in (\ref{fx}) implies $M=\tfrac{1}{2}F_z/U_z$, 
which means that (\ref{up}) integrates to yield 
$$ 
\tfrac{1}{4}(F-c)^2+kU=\ell, 
$$ 
for another integration constant $\ell$, 
which corresponds 
to an ODE of first order for $U(z)$, namely 
\beq\label{uz}
U_z^2-U^2\pm2\sqrt{\ell-kU}+c=0.
\eeq 
The latter equation is easily reduced to a quadrature, 
but a more useful approach is to employ the reciprocal transformation (\ref{mchrt}), which leads to an explicit 
parametric form for the general solution.

If we take the reciprocally transformed equation (\ref{vT}), written in the form 
$$ 
(m^{-1})_T=\tfrac{1}{8}f_X,
$$ 
then reducing to travelling waves with velocity $\tilde{c}$ we have dependent variables 
$U(Z)$, $M(Z)$, $F(Z)$, considered as functions 
of 
\beq\label{Zvar}
Z=X-\tilde{c}T,
\eeq 
and the conservation law (\ref{vT}) reduces to a total $Z$ derivative, which integrates to give 
$$-\tilde{c}M^{-1}=\tfrac{1}{8}F+\mathrm{const}. $$
The  above equation  
is equivalent to (\ref{up}) if we identify the integration constant with  $-\tfrac{1}{8}c$, 
and $k=8\tilde{c}\neq 0$, 
so that 
\beq\label{nmf} 
M(F-c)+8\tilde{c}=0,
\eeq 

Thus we see that the travelling wave reduction of the mCH equation (\ref{mchm}) corresponds to the 
travelling wave reduction of the PDE (i.e., the first negative mKdV flow) that is obtained from it via the reciprocal transformation 
(\ref{mchrt}), provided that the parameters $c,\tilde{c}$ are appropriately identified as velocities/integration constants, with 
their roles interchanged in passing between the two equations. 
Furthermore, it turns out that (\ref{mchrt}) reduces to a hodograph transformation between the ODEs obtained 
from these reductions, since 
\beq\label{hodZ} 
\begin{array}{rcl}
\rd Z & = & \rd X - \tilde{c}\,\rd T 
\\
& = & \sfrac{1}{2}m\, \rd x -\sfrac{1}{2}fm\, \rd t-4\tilde{c}\,\rd t 
\\ 
& = & \sfrac{1}{2}M\, \rd x -  \sfrac{1}{2}MF\, \rd t +  \sfrac{1}{2}M(F-c)\, \rd t 
\\
& = & \sfrac{1}{2}M\, (\rd x-c\,\rd t) \\
& = & \sfrac{1}{2}M\, \rd z.
\end{array}
\eeq 
 
There are two ways to make use of the equation (\ref{nmf}), viewed as the travelling wave reduction of the 
reciprocally transformed conservation law (\ref{vT}). First of all, the reductions of (\ref{fx}) and (\ref{mom}), transformed into 
expressions involving $Z$ derivatives, give 
\beq\label{FUZ} 
F_Z=2MU_Z, 
\eeq 
and 
$$ 
M=U-\tfrac{1}{4}M\frac{\rd}{\rd Z}(MU_Z), 
$$ 
while (\ref{fdef}) becomes 
$$ 
F=U^2-\tfrac{1}{4}(MU_Z)^2.
$$ 
Using (\ref{FUZ}) to eliminate $U_Z$ from the latter two equations, we find that 
\beq\label{UF}
U=M\big(\tfrac{1}{8}F_{ZZ}+1\big)=-\frac{8\tilde{c}}{F-c}\left(\frac{1}{8}F_{ZZ}+1\right),  
\eeq 
where (\ref{nmf}) was used to obtain the last equality, by substituting for $M$, and also 
\beq\label{UsqF}
U^2=\tfrac{1}{16}F_{Z}^2+F.   
\eeq 
Upon comparing the two expressions (\ref{UF}) and (\ref{UsqF}) for $U$, an ODE of second order and second degree 
for $F=F(Z)$ results, namely 
\beq\label{F2nd} 
\frac{64\tilde{c}^2}{(F-c)^2}\big(\tfrac{1}{8}F_{ZZ}+1\big)^2=\tfrac{1}{16}F_{Z}^2+F.  
\eeq 
However, a second way to view this reduction is to consider the quantity $\tilde{v}(Z)$ obtained by reducing 
$v(X,T)$ to a travelling wave, so that 
\beq\label{vtil} 
\tilde{v}(Z)=\frac{1}{M(Z)}=\frac{c-F(Z)}{8\tilde{c}}.
\eeq 
Applying the travelling wave reduction directly to the PDE (\ref{veqn}), it is clear that each side is a total $Z$ 
derivative, so upon integrating and rearranging, an ODE of second order for $\tilde{v}$ arises, that is 
\beq\label{vtrav} 
\tilde{c}\frac{\rd^2\tilde{v}}{\rd Z^2}-2\tilde{c}\tilde{v}^3+\tilde{k}\tilde{v}-1=0, 
\eeq 
where $\tilde{k}$ is an integration constant.

The equation (\ref{vtrav}) is solved in terms of elliptic functions. A shortcut to deriving the explicit form of these solutions is provided by Proposition \ref{miura}: there is a Miura map between the solutions of (\ref{vtrav}) and the 
travelling wave solutions of (\ref{p3rd}), as presented in \cite{bh} (see also \cite{hach}). Identifying $\tilde{c}$ with 
the wave velocity $d$ in \cite{bh}, the travelling waves of (\ref{p3rd}) correspond to a KdV field $V(Z)=-2\wp(Z)-\wp(W)$ 
(up to the freedom to replace $Z\to Z+\,$const), 
where $W$ is an arbitrary constant, so that the Miura formula (\ref{Vv}) requires that 
\beq\label{vlame}
\frac{\rd\tilde{v}}{\rd Z}-\tilde{v}^2=-2\wp(Z)-\wp(W).
\eeq 
This implies that $\tilde{v}=-\tfrac{\rd}{\rd Z}\log \psi$, where $\psi$ satisfies the Schr\"odinger equation 
\beq\label{schro}
\frac{\rd^2\psi}{\rd Z^2}-2 \wp(Z)\psi=\wp(W)\psi, 
\eeq 
equivalent to the simplest case of Lam\'e's equation. A direct calculation then shows that taking 
\beq\label{psisol}
\psi (Z) = \frac{\si (Z+W)}{\si(Z)}\exp\big(-\ze (W) Z\big),  
\eeq  
with $\si$ and $\zeta$ denoting the Weierstrass sigma and zeta functions, respectively,  
gives the general solution of (\ref{vtrav}) in the form 
\beq\label{vsol}
\tilde{v}(Z)=-\frac{1}{2}\left(\frac{\wp'(Z)-\wp'(W)}{\wp(Z)-\wp(W)}\right),
\eeq 
up to the freedom to shift $Z\to Z-Z_0$, 
for an arbitrary constant $Z_0$ (since the ODE for $\tilde{v}$ is autonomous), 
provided that the parameters are given by 
\beq\label{mchparams} 
\tilde{c}=-\frac{1}{2\wp'(W)}, \qquad \tilde{k}=6\tilde{c}\wp(W). 
\eeq

The solution (\ref{vsol}) can also be obtained more directly from the travelling wave reduction of  (\ref{p3rd}), which is given by 
$p(X,T)=P(Z)$ with 
$$ 
P(Z)=\frac{\wp(Z)-\wp(W)}{\wp'(W)} 
$$ 
(cf.\ equation (2.14) in \cite{bh}), by applying this reduction to the formula (\ref{vpinv}). This gives 
$$
\tilde{v}=-\frac{\Big(\tfrac{\rd P}{\rd Z}-1\Big)}{2P},
$$
and then substituting the explicit form of $P$ as above yields the required expression for $\tilde{v}$.  
Observe that the resulting solution (\ref{vsol}) depends on three parameters, namely $W$ and the invariants $g_2,g_3$ of
the Weierstrass $\wp$ function, so together with the arbitrary shift $Z_0$ this makes a total of four free parameters, 
corresponding to the two coefficients $\tilde{c},\tilde{k}$ plus two initial data required to specify the initial value problem for 
(\ref{vtrav}). 

The relation (\ref{vtil}) shows that the solution of the second degree equation (\ref{F2nd}) for $F$ should be given 
by 
\beq \label{Fv}
F(Z)=c-8\tilde{c}\tilde{v}(Z), 
\eeq 
with $\tilde{v}$ specified according to (\ref{vsol}), and a direct calculation shows that indeed this is the case, provided 
that the parameter $c$ is taken as 
\beq\label{cmch}
c=\frac{2\wp''(W)}{\wp'(W)^2}.  
\eeq  
Finally, comparing (\ref{UF}) with (\ref{Fv}) and (\ref{vtrav}), 
we find that the quantity $U$ can be expressed in terms of $\tilde{v}$ as 
\beq\label{Uv} 
U(Z)=-2\tilde{c}\tilde{v}(Z)^2+\tilde{k}. 
\eeq 
This allows the travelling wave solutions of the mCH equation to be expressed in parametric form.

\begin{thm}\label{mchtrav} 
The smooth travelling wave solutions (\ref{travwaves}) of the mCH equation (\ref{mch}) 
are given parametrically by $U=U(Z)$, $z=z(Z)$, where $U(Z)$ is defined by 
(\ref{Uv}) with $\tilde{v}(Z)$ as in (\ref{vsol}) (up to the freedom to shift 
$Z\to Z+\,$const), and 
\beq\label{zmch}
z(Z)=2\log\si(Z)-2\log\si(Z+W)+2\ze (W)Z+\mathrm{const},
\eeq 
with the parameters being specified by (\ref{mchparams}) and (\ref{cmch}). 
\end{thm} 
\begin{prf} 
The formula for $U(Z)$ has already been derived above, so it remains to calculate 
$z(Z)$. From (\ref{hodZ}) it follows that  
$\rd z =2M(Z)^{-1}\rd Z=2\tilde{v}(Z)\,\rd Z$, and using 
$\tilde{v}=-\sfrac{\rd}{\rd Z}\log\psi$ with $\psi$ as in (\ref{psisol}), the 
formula (\ref{zmch}) follows.  
\end{prf}

\begin{figure}
\centering 
\begin{subfigure}{.5\textwidth}
 \centering
 \includegraphics[width=.7\linewidth]{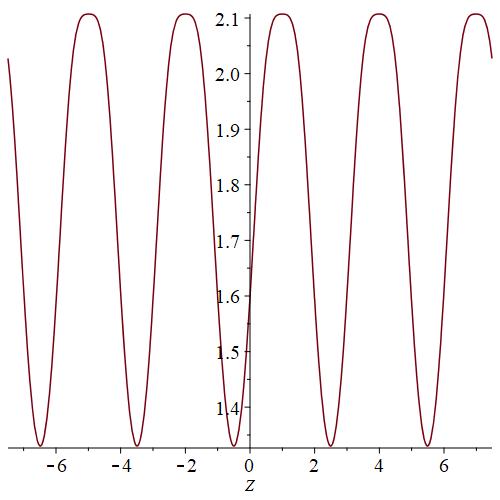}
 \caption{$U(Z)$ for $-5\om_1\leq Z\leq 5\om_1$.}\label{UvsZ} 
\end{subfigure}%
\begin{subfigure}{.5\textwidth}
 \centering
 \includegraphics[width=.7\linewidth]{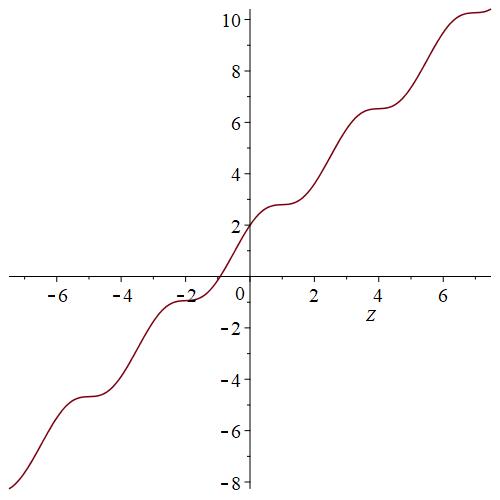}
 \caption{$z(Z)$ for $-5\om_1\leq Z\leq 5\om_1$.}\label{zvsZ} 
\end{subfigure}\caption{Parametric form of mCH travelling waves for real $Z$.}\label{Uztravplot}
\end{figure}

\begin{exa}\label{mchtravplot}
To illustrate the preceding theorem, we use it 
to plot a particular travelling wave solution of (\ref{mch}) which is bounded and real for $x,t\in\R$.
We choose a Weierstrass cubic defined by fixing the values of the invariants, and also make a choice of the 
parameter $W$, taking  
$$ 
g_2=4, \qquad g_3=-1, \qquad W=1.
$$ 
From  (\ref{cmch}) and  (\ref{mchparams}) this gives the value of the velocity of the travelling wave and 
the other constants appearing in the solution as 
$$ 
c\approx 4.494929942, \qquad \tilde{c}\approx 0.298653316, \qquad \tilde{k}\approx 2.107492133.
$$ 
In this case, the Weierstrass $\wp$ function has real/imaginary half-periods given by 
$$ 
\om_1\approx 1.496729323 
, 
\qquad 
\om_2\approx 1.225694691 \ri 
, 
$$
respectively, so taking the third half-period $\om_3=\om_1+\om_2$, 
the function $\wp(Z+\om_3)$ is real-valued, bounded and periodic with real period $2\om_1$ for $Z\in\R$.
Thus, to avoid poles for real values of $Z$, we can exploit the freedom to shift $Z$ and $z$ in Theorem \ref{mchtrav}, 
replacing $Z\to Z+\om_3$ in (\ref{vsol}) and (\ref{zmch}), and choosing the arbitrary constant in the latter so that $z$ is real for all 
$Z\in\R$. This guarantees that $U$ given by (\ref{Uv}) is a bounded periodic function for real argument $Z$, 
and the corresponding function $U(z)$ defined parametrically by $z(Z)$ is a bounded periodic solution of (\ref{uz}). Indeed, 
from the quasiperiodicity 
of the Weierstrass sigma function, which in particular means that $\si(Z+2\om_1)=-e^{2(Z+\om_1)\ze(\om_1)}\si(Z)$, it follows 
from (\ref{zmch}) that the period of $U(z)$ is given by 
$$
z(Z+2\om_1)-z(Z)=4\big(\om_1\ze(W)-W\ze(\om_1)\big)
\approx3.734925095$$ in this particular numerical example. 
\begin{figure}
 \centering
\label{Uztrw}
\epsfig{file=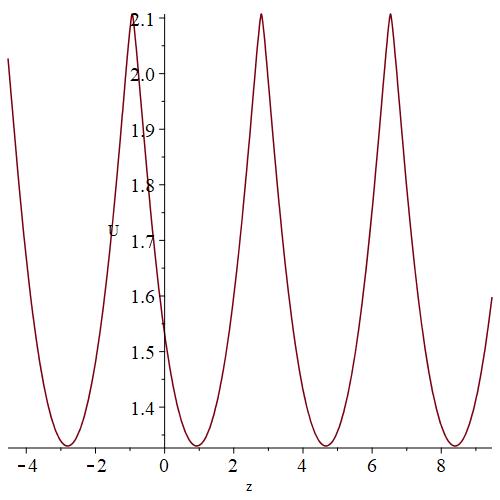, height=2.7in, width=2.7in}
\caption{Periodic travelling wave solution of the mCH equation.
}
\end{figure}%
Moreover, in this 
case we find that $\tilde{v}(Z)$ is positive for all real $Z$, so $\sfrac{\rd z}{\rd Z}=2\tilde{v}>0$ 
and $z(Z)$ is a monotone increasing function of its argument, as is visible from the right-hand panel of Fig.1. 
We have also plotted $U$ against $Z$ in the left-hand panel of the latter figure, where both plots are for 
$-5\om_1\leq Z\leq 5\om_1$, while in Fig.2 
we have plotted $U$ against $z$, in the range $-3\om_1\leq Z\leq 4\om_1$, corresponding to the travelling wave profile for the mCH equation. 
Although the periodic peaks in this figure appear somewhat sharp, a closer look reveals that the solution is smooth for 
all real $z$.  
However, by suitably adapting the technique used in \cite{lo}, it should be possible to obtain a single  
(weak) peakon solution from this family of periodic solutions, by taking a double scaling limit where the real 
period $2\om_1\to\infty$ and the background (minimum) value of $U$ tends to zero. 
\end{exa}

\subsection{Scaling similarity solutions}

The mCH equation (\ref{mch}) has a one-parameter family of similarity solutions given by taking 
\beq\label{simred}
u(x,t)=t^{-\sfrac{1}{2}}U(z), \qquad m(x,t)=t^{-\sfrac{1}{2}}M(z), \qquad z=x+\al\log t.
\eeq 
These solutions generalize the separable solutions (\ref{sep}), 
which arise when the parameter $\al=0$. Upon substituting the expressions (\ref{simred}) 
into the mCH equation (\ref{mch}), or equivalently into (\ref{mchm}), we find 
\beq\label{UMeq}
(\al+U^2-U_z^2)M_z+(2U_zM-\sfrac{1}{2})M=0, \qquad 
M=U-U_{zz}, 
\eeq
where the latter is a compact way of writing the corresponding autonomous ODE 
of third order for $U(z)$, that is 
\beq\label{Usimz} 
(U_z^2-U^2-\al)(U_{zzz}-U_z)+2U_z(U_{zz}-U)^2+\sfrac{1}{2}(U_{zz}-U)=0. 
\eeq 

In order to obtain parametric formulae for the solutions of (\ref{Usimz}), 
we proceed to apply the reciprocal transformation (\ref{mchrt}) to 
the similarity solutions (\ref{simred}). Without loss of generality we can fix 
\beq\label{fixt}
T=4t\implies \rd \log T=\rd \log t,
\eeq
and then we find that, under the reciprocal transformation,  the reductions (\ref{simred}) 
correspond to scaling similarity reductions of the first negative mKdV flow, 
obtained by taking 
\beq\label{rtsimred} 
u=2T^{-\sfrac{1}{2}}U(Z), \quad 
m=2T^{-\sfrac{1}{2}}M(Z), \quad
f=4T^{-1}F(Z), \quad
Z=XT^{\sfrac{1}{2}}.
\eeq  
Indeed, applying the reduction (\ref{rtsimred}) directly to the conservation law (\ref{vT}) 
with $v=m^{-1}$ produces 
$$ 
\frac{\rd }{\rd Z}\big(\sfrac{1}{2}ZM^{-1}\big)=\frac{\rd F}{\rd Z}\implies 
\sfrac{1}{2}ZM^{-1}-F=\mathrm{const}, 
$$ 
and if we identify the integration constant above with the parameter $\al$ then this yields the 
relation 
\beq\label{alrel}
M(F+\al)=\sfrac{1}{2}Z. 
\eeq 
To see that this is consistent with applying the reciprocal transformation to the 
solutions (\ref{simred}), note that from (\ref{alrel}) and the relation (\ref{fixt}) 
between $t$ and $T$ we have 
\beq\label{Zcalc}
\begin{array}{rcl} 
\rd Z & = & T^{\sfrac{1}{2}}\rd X+\sfrac{1}{2} T^{-\sfrac{1}{2}}X\rd T \\ 
&=&  T^{\sfrac{1}{2}}\rd X + \sfrac{1}{2}Z\, \rd \log T \\
& = & T^{\sfrac{1}{2}}(\sfrac{1}{2}m\,\rd x-\sfrac{1}{2}mf\, \rd t) 
+M(F+\al)\,\rd \log t \\
&=& T^{\sfrac{1}{2}}(T^{-\sfrac{1}{2}}M\,\rd x -4T^{-\sfrac{3}{2}}MF\,\rd t)
+M(F+\al)\,\rd \log t \\
&=& M(\rd x+\al\,\rd \log t) \\
& = & M\,\rd z.
\end{array}
\eeq
As shown previously, the reduction (\ref{simred}) produces the ODE (\ref{UMeq}) 
with independent variable $z$, 
which  can be rewritten  in terms of $M$ and $F=U^2-U_z^2$ in the form 
$$ 
\frac{\rd}{\rd z}\Big(M(F+\al)\Big)=\sfrac{1}{2}M. 
$$ 
Then using the result of (\ref{Zcalc}) to transform the derivatives according to 
$\sfrac{\rd}{\rd z}=M\sfrac{\rd}{\rd Z}$, the latter ODE becomes 
$$ 
\frac{\rd}{\rd Z}\Big(M(F+\al)\Big)=\frac{1}{2}, 
$$
that is precisely the outcome of differentiating each side of (\ref{alrel}) 
with respect to $Z$. 

Now from the definitions (\ref{fdef}) and (\ref{mom}), we can write down corresponding 
relations for the similarity solutions (\ref{rtsimred}), involving $Z$ derivatives, namely 
$$
F=U^2-\left(M\frac{\rd U}{\rd Z}\right)^2, 
\qquad 
M= U-\left(M\frac{\rd }{\rd Z}\right)^2 U, 
$$ 
and then from (\ref{fx}) we obtain 
\beq\label{FZUZ}
\frac{\rd F}{\rd Z}=2M \frac{\rd U}{\rd Z}, 
\eeq
so we can eliminate derivatives of $U$ from the previous identities, to find 
\beq\label{UFeqs} 
U^2=\tfrac{1}{4}F_Z^2+F, \qquad 
U=M\big(\sfrac{1}{2}F_{ZZ}+1\big).
\eeq 
Upon comparing the two equations for $U$ above, and using (\ref{alrel}) to 
write $M$ in terms of $F$, we obtain a single ODE of second order and second degree 
satisfied by $F$, that is 
\beq\label{F2nddeg}
\left(\frac{1}{2}\frac{\rd^2 F}{\rd Z^2}+1\right)^2=
\frac{(F+\al)^2}{Z^2}
\left(\Big(\frac{\rd F}{\rd Z}\Big)^2+4F\right), 
\eeq 
which is a non-autonomous analogue of (\ref{F2nd}). 

If we introduce $\tilde{v}=\tilde{v}(Z)$ according to 
\beq\label{tvdef}
\tilde{v}=M^{-1}=2Z^{-1}(F+\al),
\eeq
so that 
\beq\label{vscale} 
v(X,T)=\tfrac{1}{2}T^{\sfrac{1}{2}}\tilde{v}(Z),
\eeq  
then the direct similarity reduction of (\ref{veqn}) is an 
ODE of third order for $\tilde{v}$, given by 
\beq\label{tv3rd} 
\frac{\rd}{\rd Z} \left(\tilde{v}^{-1}\Big(\frac{\rd^2}{\rd Z^2}(Z \tilde{v})+4  
\Big)\right) 
=\tilde{v}\frac{\rd}{\rd Z} \big(Z\tilde{v}\big).
\eeq 
However, unlike the case of travelling waves, the above equation cannot be integrated 
to yield an analogue of (\ref{vtrav}) that is first degree in $\tilde{v}_{ZZ}$; instead, 
$\tilde{v}$ satisfies a non-autonomous equation of second order and second degree, obtained from (\ref{F2nddeg}) by replacing 
$F$ and its derivatives, using  
\beq\label{Fvsub} 
F=\tfrac{1}{2}Z\tilde{v}-\al, 
\eeq 
which follows from (\ref{tvdef}). 

We now proceed to show how $F(Z)$ satisfying (\ref{F2nddeg}), or equivalently $\tilde{v}(Z)$, is given 
in terms of a solution of a particular case of Painlev\'e III, that is 
\beq\label{p3} 
\frac{\rd^2 P}{\rd Z^2}=\frac{1}{P}\left(\frac{\rd P}{\rd Z}\right)^2-\frac{1}{Z}\left(\frac{\rd P}{\rd Z}\right)+\frac{1}{Z}\big( 2P^2+a) -\frac{1}{P}, 
\eeq
for a suitable choice of the parameter $a$. If we identify $P\to w$ and $Z\to \ze$, then this is 
equation (\ref{piii}) with parameters
$$
\tilde{\al}=2, 
\quad 
 \tilde{\be}=a, 
\quad 
\tilde{\gam}=0, \quad 
\tilde{\delta}=-1
.
$$ 
The main point is that, as described in \cite{bh} (and originally derived in \cite{hach}), the ODE (\ref{p3}) arises as the equation for 
scaling similarity reductions of the negative KdV flow (\ref{p3rd}), so by applying the result of Proposition \ref{miura} to these reductions, a link with (\ref{F2nddeg}) 
follows. Under the reduction (\ref{simred}) applied to (\ref{pv}) with $v(X,T)=\sfrac{1}{2}T^{\sfrac{1}{2}}M^{-1}=\sfrac{1}{2}T^{\sfrac{1}{2}}\tilde{v}$, we 
find 
\beq\label{pred}
p(X,T)=T^{-\sfrac{1}{2}}P(Z),  
\eeq 
where $P=P(Z)$ is given in terms of $\tilde{v}$ by
\beq\label{Pvdef} 
P=\frac{1}{\tilde{v}}\left(\frac{1}{4}\frac{\rd^2}{\rd Z^2}\big( Z\tilde{v}\big) +1\right)-\frac{1}{4}\frac{\rd}{\rd Z}\big(Z\tilde{v}\big), 
\eeq 
or equivalently, using (\ref{tvdef}), it can be rewritten in terms of $F$ as 
\beq\label{PFdef}
P=\frac{Z}{2(F+\al)}\left(\frac{1}{2}\frac{\rd^2 F}{\rd Z^2}+1\right) - \frac{1}{2}\frac{\rd F}{\rd Z}.
\eeq 
Since we know from \cite{bh} that if $p(X,T)$ satisfying (\ref{p3rd}) has the form (\ref{pred}) then 
$P(Z)$ is a solution of (\ref{p3}) for some $a$, the question is how to determine this parameter. 
It is convenient to note that (\ref{Pvdef}) or (\ref{PFdef}) can be expressed in the form 
\beq\label{PU}
P=U-\tfrac{1}{2}F_Z,  
\eeq 
and also observe that (\ref{FZUZ}) is equivalent to 
the formula 
$$ 
U_Z=\tfrac{1}{2}\tilde{v}F_Z.
$$ 
Then applying $\sfrac{\rd}{\rd Z}$ to (\ref{PU}) together with the second equation in (\ref{UFeqs})  implies that 
$$ 
P_Z=U_Z-\tfrac{1}{2}F_{ZZ}=\tfrac{1}{2}\tilde{v}F_Z+1-\tilde{v}U\implies 
P_Z-1=\tilde{v}\big(\tfrac{1}{2}F_Z-U\big)=-\tilde{v}P.
$$ 
Hence we obtain the following expression for $\tilde{v}$ in terms of $P$:
\beq\label{vPform}
\tilde{v}=-\frac{(P_Z-1)}{P}. 
\eeq 
This also follows directly by applying the similarity reduction to the formula (\ref{vpinv}), taking the scaling 
(\ref{vscale}) into account, 
and it leads to the relation between the solutions of  (\ref{F2nddeg}) and (\ref{p3}). 

\begin{lem}\label{corr}
There is a one-to-one correspondence between solutions of  (\ref{F2nddeg}) and (\ref{p3}), with the parameters related by 
\beq\label{aalpha} 
a=2\al +1, 
\eeq where $P$ is given in terms of $F$ by (\ref{PFdef}), and $F$ is given in terms of $P$ by 
\beq\label{FPdef}
F=-\frac{Z}{2P}\left(\frac{\rd P}{\rd Z}-1\right)-\al. 
\eeq 
\end{lem}
\begin{prf}
The relation (\ref{FPdef}) for $F$ is an immediate consequence of (\ref{vPform}) and (\ref{Fvsub}), and can be rewritten as 
$$ 
F=-\tfrac{1}{2}Z\La_Z+\tfrac{Z}{2}P^{-1}-\al, \qquad \mathrm{with}\quad \La_Z=\tfrac{\rd}{\rd Z}\log P, 
$$ 
where the latter notation allows the Painlev\'e III equation (\ref{p3}) to be expressed as 
$$
\frac{\rd}{\rd Z} \big(Z\La_Z\big)=2P+aP^{-1}-ZP^{-2}.  
$$
Using this form of the ODE for $P$ to eliminate  terms in $\La_{ZZ}=\sfrac{\rd^2}{\rd Z^2}\log P$, the derivatives of $F$ can be written as
$$ 
F_Z=-\tfrac{Z}{2}P^{-1}\La_Z-P+\tfrac{1}{2}(1-a)P^{-1}+\tfrac{Z}{2}P^{-2},
$$ 
$$
F_{ZZ}=\tfrac{Z}{2}P^{-1}\La_Z^2-\big(P+\tfrac{1}{2}(1-a)P^{-1}+ZP^{-2}\big)\La_Z-1+\tfrac{1}{2}(1-a)P^{-2}+\tfrac{Z}{2}P^{-3}.
$$
Upon substituting these expressions for $F$ and its derivatives into (\ref{F2nddeg}), almost all the terms cancel, and all that remains 
is 
$$
\frac{1}{2}\big(\La_Z-P^{-1}\big)^2(2\al+1-a)=0, 
$$
from which (\ref{aalpha}) follows.
\end{prf}

In the description of the solutions of (\ref{Usimz}) in parametric form, it is convenient to make use of 
solutions of the ODE (\ref{p3}) connected via a B\"acklund transformation. For this case of 
the Painlev\'e III equation, given a solution $P=P(Z)$ with parameter value $a$, the quantities 
\beq\label{Ppm}
P_\pm = \frac{Z\big(\pm P_Z+1\big)}{2P^2} + \frac{\big(\mp 1-a\big)}{2P} 
\eeq 
are solutions of the same equation but with the parameter replaced by $a\pm 2$, respectively. 
It is also helpful to consider the form of the corresponding KdV field $V$ under the 
scaling similarity reduction (\ref{pred}), 
which takes the form 
\beq\label{Vred} 
V(X,T)=T\bar{V}(Z), \qquad \bar{V}=-\frac{1}{4P^2}\left(\Big(\frac{\rd P}{\rd Z}\Big)^2-1\right) 
+ \frac{1}{2ZP}\left( 
\frac{\rd P}{\rd Z}-2P^2-a
\right) ,  
\eeq 
where the above expression for $\bar{V}$ in terms of $P$ is found by applying the 
similarity reduction to the formula (\ref{ep}), and then using (\ref{p3}) to 
eliminate the $P_{ZZ}$ term. Then we introduce a tau function $\si_a (Z)$, 
in terms of which the scaled KdV field $\bar{V}$ is given by the standard KdV tau function relation 
\beq\label{kdvtau}\bar{V}(Z)=2\frac{\rd^2}{\rd Z^2}\log \si_a (Z)\eeq  
(invariant under gauge transformations of 
the form $\si_a (Z)\to \exp (AZ+B)\si_a(Z)$). The index $a$ denotes the parameter value in the equation (\ref{p3}), so 
if we replace $P\to P_{\pm}$ and $a\to a\pm 2$ in the formula (\ref{Vred}) for $\bar{V}$ then we obtain corresponding 
(scaled) KdV fields $\bar{V}_\pm$ and their associated tau functions, related by 
$$ \bar{V}_\pm(Z) = 2\frac{\rd^2}{\rd Z^2}\log \si_{a\pm 2} (Z). $$ 

\begin{remark}
In addition to a fixed singularity at $Z=0$, where solutions can have branching (see Example \ref{Uzex} below), 
there are two kinds of movable singularities that occur in (\ref{p3}) at points $Z_0\in\C$ with $Z_0\neq 0$: 
movable zeros, where $P$ has a local expansion 
$$
P(Z)=\pm (Z-Z_0) +\frac{\pm 1-a}{2Z_0}(Z-Z_0)^2 +c_3(Z-Z_0)^3+O\big((Z-Z_0)^4\big), \quad c_3\,\,\mathrm{arbitrary}, 
$$  
and movable poles, in the neighbourhood of which $P$ has the Laurent series 
$$ 
P(Z)=Z_0(Z-Z_0)^{-2} +c_0-\frac{c_0}{Z_0}(Z-Z_0) +O\big((Z-Z_0)^2\big), \quad c_0\,\,\mathrm{arbitrary}.
$$
The reduced KdV field $\bar{V}$, given in terms of $P$ by  (\ref{Vred}), is regular at points where $P$ has movable 
zeros, since $\bar{V}(Z)=\mp\sfrac{3}{2}c_3+O\big((Z-Z_0)\big)$ in the neighbourhood of such points; 
but at points where $P$ has double poles, $\bar{V}$ does also, having the local expansion
$$ 
\bar{V}(Z)=-2(Z-Z_0)^{-2}+\frac{c_0}{Z_0}+O\big((Z-Z_0)^2\big), 
$$ 
so from (\ref{kdvtau}) the tau function vanishes at these points, being given 
by 
$$ 
\si_a(Z)=C(Z-Z_0)\Big(1+O\big((Z-Z_0)\big)\Big), \qquad C\neq 0, 
$$ 
where the constant $C$ depends on the choice of gauge.
\end{remark}

Using loop group methods, Schiff constructed a B\"acklund transformation for the PDE (\ref{p3rd}) \cite{schiff}, 
and in \cite{hach} it was remarked that this arises naturally from the standard Darboux transformation for the 
Schr\"odinger operator. Furthermore, in the case of the scaling similarity solutions of (\ref{p3rd})   it corresponds to a 
Darboux transformation with zero eigenvalue, 
which 
is associated with 
the operator refactorization 
$$ 
D_X^2+V=(D_X-v)(D_X+v) \longrightarrow 
D_X^2+V^*=(D_X+v)(D_X-v), 
$$ 
and this reduces to the B\"acklund transformation (\ref{Ppm}) for the Painlev\'e III equation (\ref{p3}). 
In terms of the standard Miura formula  (\ref{Vv}), the latter transformation is achieved by replacing $v\to -v$; but more 
precisely, at the level of the scaling similarity reduction, taking into account the factors of 2 that appear in (\ref{rtsimred}), a 
direct calculation shows that this transformation gives 
\beq\label{VvBT} 
\bar{V}=\tfrac{1}{2}\tilde{v}_Z-\tfrac{1}{4}\tilde{v}^2\longrightarrow 
\bar{V}_-=-\tfrac{1}{2}\tilde{v}_Z-\tfrac{1}{4}\tilde{v}^2, \qquad a\to a-2.
\eeq 
Subtracting the expressions for $\bar{V}$ and $\bar{V}_-$ produces 
$$ 
\tilde{v}_Z=\bar{V} - \bar{V}_-
, 
$$ 
which leads to the usual expression for an mKdV field as the logarithmic derivative of a ratio of 
two tau functions: the scaled field $\tilde{v}$ is given by 
\beq\label{vtau}
\tilde{v}(Z)=2\frac{\rd}{\rd Z} \log \left(\frac{\si_{a} (Z)}{\si_{a- 2} (Z)}\right) 
\eeq 
(with an appropriate choice of gauge). 
This allows us to state the main result of this section. 
\begin{thm}\label{mchmain} 
The solutions of the ODE (\ref{Usimz}) for the similarity reduction (\ref{simred}) of the mCH equation (\ref{mch}) 
are given parametrically by $U=U(Z)$, $z=z(Z)$, where $U$ is defined by 
\beq\label{UFform}
U=\frac{Z}{2(F+\al)}\left(\frac{1}{2}\frac{\rd^2 F}{\rd Z^2}+1\right), 
\eeq
with $F(Z)$ being a solution of the ODE  (\ref{F2nddeg}), related to a solution 
of the Painlev\'e III equation (\ref{p3}) with parameter $a=2\al+1$ by  (\ref{FPdef}), and 
\beq\label{zsimmch}
z(Z)=2\log\si_a(Z)-2\log\si_{a-2}(Z)+\mathrm{const},
\eeq 
in terms of two Painlev\'e III tau functions $\si_a$, $\si_{a-2}$ connected via a B\"acklund transformation. 
\end{thm}

\begin{prf}
The formula (\ref{UFform}) follows from the second relation in (\ref{UFeqs}) together with (\ref{tvdef}), while 
(\ref{zsimmch}) comes from rearranging (\ref{Zcalc}) as $\rd z = M^{-1}\rd Z=\tilde{v}\,\rd Z$ 
and using (\ref{vtau}) to  integrate this. 
\end{prf}

\begin{figure}
 \centering
\label{Uzfig}
\epsfig{file=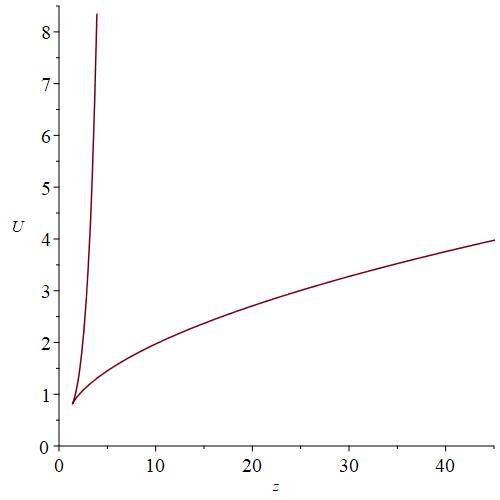, height=2.7in, width=2.7in}
\caption{Plot of $U$ against $z$ for the parametric solution 
(\ref{Uzparam}) with $0.02\leq \ze\leq 4$.
}
\end{figure} 

\begin{exa}\label{Uzex} 
It is instructive to consider an explicit example 
of the parametrization in Theorem \ref{mchmain}. 
The equation (\ref{p3}) has a family of algebraic solutions 
for even integer values of the parameter $a$ (see \cite{clarkson} 
and references). 
The simplest such solution is given by 
\beq\label{P0}
P=(Z/2)^{\sfrac{1}{3}}, \qquad a=0. 
\eeq  
It is convenient to write all formulae in terms of $\ze=(Z/2)^{\sfrac{1}{3}}$, so that 
from (\ref{Vred}) we have 
\beq\label{Vbar0}
P=\ze \implies \bar{V}=\tfrac{5}{144}\ze^{-6}-\tfrac{1}{4}\ze^{-2}.
\eeq 
The tau functions for some of these algebraic solutions are listed in Table 1 of \cite{bh}, the 
relevant ones here being 
\beq\label{0m2taus} 
\si_0 =\ze^{-\tfrac{5}{24}}\exp \big(-\tfrac{9}{8}\ze^4\big), 
\qquad 
\si_{-2} =\ze^{\tfrac{7}{24}}\exp \big(-\tfrac{9}{8}\ze^4-\tfrac{3}{2}\ze^2\big).
\eeq  
These generate the reduced mKdV field according to  
$$ 
\tilde{v}=\frac{\rd}{\rd Z}\log\left(\frac{\si_0}{\si_{-2}}\right)
=\frac{1}{6}\ze^{-2}\frac{\rd}{\rd \ze}\log\left(\frac{\si_0}{\si_{-2}}\right)=-\frac{1}{6}\ze^{-3}+\ze^{-1}, 
$$ 
so that the reduced KdV field  as in (\ref{Vbar0}) is given by 
$$ 
\bar{V}=\tfrac{1}{2}\tilde{v}_Z-\tfrac{1}{4}\tilde{v}^2=\tfrac{1}{12}\ze^{-2}\tilde{v}_\ze-\tfrac{1}{4}\tilde{v}^2,
$$
and from (\ref{Fvsub}) the corresponding solution of  (\ref{F2nddeg}) is found to be 
\beq\label{Fsol}
F=\ze^2+\frac{1}{3}, \qquad \al =-\frac{1}{2}.
\eeq 
Then from (\ref{UFform}) and (\ref{zsimmch}), using the form of $F$ in (\ref{Fsol}) above and the specific tau functions (\ref{0m2taus}), 
the parametric form of the associated solution $U=U(z)$ of the ODE 
 (\ref{Usimz}) with $\al=-\sfrac{1}{2}$ is 
\beq\label{Uzparam}
U=\ze +\tfrac{1}{6}\ze^{-1}, \qquad z=3\ze^2-\log \ze,  
\eeq 
where here the solution is parametrized by $\ze$ instead of $Z$, and a choice of arbitrary constant in $z$ has been set to zero. 
If we consider this solution for real $\ze>0$, then it is clear  
that this  similarity reduction 
of the mCH equation (\ref{mch}) has (at least) two real branches: 
in the limit of small positive $\ze$, we have 
$$ 
\ze\to 0+\implies z\to\infty, \quad U\to\infty, 
$$ 
with asymptotics 
\beq\label{expasy}
U\sim\frac{1}{6}\ze^{-1}, \quad z\sim -\log \ze \implies U\sim \frac{1}{6}e^z, 
\eeq 
while for large $\ze$, we find 
$$ 
\ze\to \infty\implies z\to\infty, \quad U\to\infty, 
$$ 
but the asymptotic behaviour is completely different, namely 
$$ 
U\sim\ze, \quad z\sim 3\ze^2 \implies U\sim \sqrt{\frac{z}{3}}. 
$$ 
From the tangent vector 
$$ 
\left(\begin{array}{c} 
\sfrac{\rd z}{\rd \ze} \\ 
 \sfrac{\rd U}{\rd \ze}
\end{array}
 \right) 
=
\left(\begin{array}{c} 
6\ze-\ze^{-1} \\ 
 1-\sfrac{1}{6}\ze^{-2}
\end{array}
 \right) 
$$
we see that 
$\sfrac{\rd z}{\rd \ze} =0= 
 \sfrac{\rd U}{\rd \ze}$ 
when $\ze=1/\sqrt{6}$, corresponding 
to 
$(z,U)=\big(\sfrac{1}{2}(1+\log 6),2/\sqrt{6}\big)$, where  
the two branches of the solution separate at a cusp, 
clearly visible in Fig.3. 
Note that the exponential asymptotics in (\ref{expasy})  can be regarded as  the leading term 
in an expansion 
$U\sim \sfrac{1}{6}e^z+\sum_{n\geq 0} c_ne^{-nz}$ as $z\to\infty$, which is 
the sort of exponential series considered for Camassa-Holm type equations in \cite{pickering}. 
\end{exa}

\section{Reductions of Novikov's equation} 
 
\setcounter{equation}{0}

This section is devoted to similarity reductions of Novikov's equation (\ref{novikov}), or equivalently (\ref{vn}). 
In due course we will need to compare its scaling similarity reductions to analogous reductions of the 
Degasperis-Procesi equation, so it will be convenient to rewrite (\ref{dp}) in the form of an associated 
conservation law for the field $p=m^{1/3}$, 
given by the system 
\beq\label{dpbar} 
\frac{\partial p}{\partial \bar{t}}+\frac{\partial }{\partial \bar{x}}(\bar{u}p)=0, \qquad 
p^3=(1-D_{\bar{x}}^2)\bar{u},
\eeq 
where the other dependent/independent variables $u,x,t$ have been replaced by $\bar{u},\bar{x},\bar{t}$. 
In the rest of this section we reserve $m,u,x,t$ for the corresponding dependent/independent variables in 
Novikov's equation (\ref{vn}). 

\subsection{Negative Kaup-Kupershmidt and Sawada-Kotera flows} 

As is well known \cite{fg}, each of 
the flows in the Kaup-Kupershmidt hierarchy (with dependent variable $V$) and in the Sawada-Kotera hierarchy (with dependent 
variable $\hat{V}$) arises as the compatibility condition of a linear system, whose $X$ part is 
given by the eigenvalue problem for a  third order Lax operator, that is  
\beq\label{kksklax}
D_X^3+4VD_X+2V_X=(D_X+v)D_X(D_X-v) \quad \mathrm{and} \quad
D_X^3+\hat{V}D_X=(D_X-v)(D_X+v)D_X, 
\eeq
respectively, where the operator factorizations above produce the Miura maps 
\beq\label{kkskmiura}
V=-\tfrac{1}{2}v_X-\tfrac{1}{4}v^2, \qquad \hat{V}=v_X-v^2, 
\eeq 
which relate each of these hierarchies to the same modified hierarchy with dependent variable $v$. 
(The choice of scale for the Kaup-Kupershmidt field $V$ in (\ref{kksklax}) differs by a factor of 2 from \cite{fg}, 
but is taken for consistency with the form of the Lax pair derived in \cite{dhh1}, and the 
results in \cite{bh}.)

In terms of the field $p$, the reciprocal transformation associated with the Degasperis-Procesi   conservation law (\ref{dpbar}) 
is identical to that for 
the Camassa-Holm case, that is 
\beq\label{dprt}
\rd X=p\,\rd \bar{x} -\bar{u}p\,\rd \bar{t},\qquad \rd T=\rd \bar{t}, 
\eeq 
apart from the fact that $p$ has a different meaning. 
The   result of 
applying this transformation (\ref{chrt}) is a PDE of third order for $p=p(X,T)$ 
as a function of the new independent variables $X,T$, given in 
conservation form as
\beq\label{dp3rd}
\frac{\partial }{\partial T} (p^{-1})  +  \frac{\partial }{\partial X}\big(p (\log p)_{XT}-p^3\big)=0.
\eeq  
The connection 
with the Kaup-Kupershmidt hierarchy is made manifest by rewriting  (\ref{dp3rd}) in the   alternative  form 
\beq\label{VTdp}
\frac{\partial V}{\partial T}+\frac{3}{4}\frac{\partial }{\partial X}\big( p^2\big)=0, 
\eeq    
where the quantity $V$ is defined in terms of $p$ by the same formula  as in (\ref{ep}) above, 
and corresponds to the dependent variable appearing in the first of the third order Lax operators in (\ref{kksklax}); 
this is how (\ref{VTdp}) was first derived in \cite{dhh1}. The PDE (\ref{VTdp}) is a flow of weight $-1$ in 
the Kaup-Kupershmidt hierarchy. 

As for Novikov's equation (\ref{vn}), if we introduce the dependent variables 
\beq\label{qr} 
q=m^{2/3}, \qquad r=um^{1/3}, 
\eeq 
then it has a conservation law with density $m^{\sfrac{2}{3}}$, which can be written as 
\beq\label{vncons} 
q_t+(r^2)_x=0, \qquad q^{2}=r-q^{\sfrac{1}{2}}\frac{\partial^2}{\partial x^2}\big( q^{-\sfrac{1}{2}} r\big), 
\eeq 
with the latter equation being (\ref{mom}) expressed in terms of $q$ and $r$. 
This conservation law is discussed in the context of the prolongation structure of the PDE in \cite{ss}. 
In order to relate this to the Sawada-Kotera 
hierarchy, it is necessary to introduce the reciprocal transformation 
\beq\label{vnrt}
\rd X=q\,\rd {x} -r^2\,\rd t,\qquad \rd T=\rd {t}, 
\eeq 
which produces the transformed system 
\beq\label{skneg} 
\frac{\partial }{\partial T} (q^{-1})  = \frac{\partial }{\partial X}\big(q^{-1}r^2\big), 
\qquad r_{XX}+\hat{V}r+1=0, 
\eeq 
where 
\beq\label{Vhatdef}
\hat{V} = -\frac{\big(\sqrt{q}\big)_{XX}}{\sqrt{q}}-\frac{1}{q^2}
\eeq 
corresponds to the dependent variable for the Sawada-Kotera hierarchy. As shown in 
\cite{hw2}, the system (\ref{skneg}) is a flow of weight $-1$ in this hierarchy, 
arising as the compatibility condition for a Lax pair whose $X$ part is the eigenvalue problem for the 
second operator in (\ref{kksklax}). 
For the discussion that follows, it will sometimes be convenient to refer to a potential $\Phi$ associated with the 
conservation law in (\ref{skneg}), so that 
\beq\label{qrphi} 
q^{-1}=\Phi_X, \qquad r=\sqrt{\frac{\Phi_T}{\Phi_X}}.
\eeq

\begin{remark}
The second Miura formula in (\ref{kkskmiura}) defines $\hat{V}$ in terms of $v$, but the solution of the inverse problem 
of finding $v$ given $\hat{V}$ is not unique, because it involves the solution of a Riccati equation. However, if 
$\hat{V}$ is specified in terms of $q$ by the formula (\ref{Vhatdef}), then taking 
\beq\label{vqform} 
v=
-\frac{1}{2}\big(\log q\big)_X\pm \frac{1}{q}
\eeq 
gives a particular solution for $v$, valid with either choice of sign above. 
\end{remark} 

\subsection{Travelling waves} 

The travelling wave solutions of the Degasperis-Procesi equation (\ref{dp}) were described 
in parametric form in \cite{bh}. They are given in terms of a parameter $Z$ which corresponds to the similarity variable 
for travelling waves 
of  the PDE (\ref{dp3rd}) with velocity $d$,  
which take the form $p(X,T)=P(Z)$, 
where $Z=X-dT$ and (up to the freedom to shift $Z\to Z+\,$const) 
\beq\label{negkktrav} 
P = \frac{1}{\al \wp'(W_1)}\left(\frac{\wp(Z)-\wp(W_1)}{\wp(Z)-\wp(W_2)}\right), \qquad d=-\frac{16\wp'(W_2)^6}{\wp'(W_1)^2\wp''(W_2)^4},
\eeq 
with constant parameters $W_1,W_2$ related via 
\beq\label{alphatrav} 
\al = -\sfrac{1}{2}\wp''(W_2)/\wp'(W_2)^2=\big(\wp(W_1)-\wp(W_2)\big)^{-1}. 
\eeq 
The corresponding velocity 
for the travelling waves of (\ref{dp}) is also given by an 
expression in terms of the Weierstrass $\wp$ function and  its derivatives with these constants as arguments; 
see \cite{bh} for details. 

The travelling waves for Novikov's equation (\ref{vn}) were reduced to a quadrature in \cite{hw2}, given as a sum of two elliptic integrals of the third kind. 
Here we derive an explicit parametric form of these travelling wave solutions, which are found by imposing the reduction 
\beq\label{vntravwaves}
u(x,t)=U(z), \qquad m(x,t)=M(z), \qquad z=x-ct  
\eeq 
in (\ref{vn}), and noting that with 
$$ 
q(x,t)=Q(z), \qquad r(x,t)=R(z), 
$$ 
the conservation law in (\ref{vncons}) integrates to yield 
$$ 
-cQ+R^2=\mathrm{const}. 
$$ 
Then, in a similar way to the case of mCH travelling waves treated in the previous section, 
we can relate these solutions with corresponding travelling waves of the reciprocally transformed 
system (\ref{skneg}) with velocity $\tilde{c}$, which we can identify (up to a sign) with the integration constant 
above, to find 
\beq\label{QRrel}
cQ=R^2+\tilde{c},  
\eeq   
in terms of $Z=X-\tilde{c}T$, with $Q=Q(Z)$ and $R=R(Z)$ related by 
\beq\label{QRode} 
\frac{R_{ZZ}+1}{R}=\frac{\big(\sqrt{Q}\big)_{ZZ}}{\sqrt{Q}}+\frac{1}{Q^2}, 
\eeq  
using the reduction of (\ref{Vhatdef}) with 
$\hat{V}\to\hat{V}(Z)$. 
To see this, note that for the travelling wave solutions of the negative Sawada-Kotera flow, the first equation in (\ref{skneg}) becomes 
$$ 
-\tilde{c}\frac{\rd}{\rd Z}\big( Q^{-1} \big)= \frac{\rd}{\rd Z}\big( Q^{-1} R^2\big),
$$
which integrates to give (\ref{QRrel}), with $c$ now playing the role of an integration constant.
Upon using (\ref{QRrel}) with the assumption $c\neq 0$ to eliminate $Q$ from (\ref{QRode}), an 
ODE of second order for $R(Z)$ is obtained, 
that is 
$$ 
\tilde{c}\left(\frac{R_{ZZ}}{R^2+\tilde{c}}-\frac{RR_Z^2}{(R^2+\tilde{c})^2}\right)
=\frac{c^2R}{(R^2+\tilde{c})^2}-1, 
$$ 
and this can be integrated to produce the first order equation 
\beq\label{Rode}
\tilde{c}\left(\frac{\rd R}{\rd Z}\right)^2 +2(R-e)(R^2+\tilde{c})+c^2=0,
\eeq 
with $e$ being another constant. 
The travelling wave reduction  of  (\ref{skneg}) corresponds to the potential in 
(\ref{qrphi})   being of the form $\Phi(X,T)=\varphi(Z)+cT$. 

Up to the freedom to shift $Z$  by an arbitrary amount $Z_0$, so that $Z\to Z-Z_0$, the general solution 
of (\ref{Rode}) for $\tilde{c}\neq 0$ is an elliptic function of $Z$ given by 
\beq\label{Rtrav}
R(Z)=-2\tilde{c}\big(\wp(Z)-\wp(W)\big),  
\eeq 
where $W$ is a constant such that 
\beq\label{vnparam}
\tilde{c}=\frac{1}{2\wp''(W)}, \qquad \frac{e}{\tilde{c}} = 6\wp(W), 
\eeq
and the other constant appearing in the ODE satisfies 
\beq\label{vnc}
\frac{c^2}{4\tilde{c}^3}=6\wp(W)\wp''(W)-\wp'(W)^2
.
\eeq 
Hence, up to shifting the argument by $Z_0$, the solution $R(Z)$ is completely 
specified by the value of the parameter $W$ and the two invariants 
$g_2,g_3$ of the Weierstrass $\wp$ function. 

By applying the reciprocal transformation (\ref{vnrt}) to the travelling waves of (\ref{vn}), and using the relation 
(\ref{QRrel}), a short calculation analogous to (\ref{hodZ}) shows that these are related to the travelling wave 
reduction of (\ref{skneg}) by the hodograph transformation 
\beq\label{vnhodZ} 
\rd Z = Q \, \rd z, 
\eeq 
where each  of the parameters $c,\tilde{c}$ has a complementary role as a wave velocity/integration constant, which switches 
according to whether the reduction of (\ref{vn}) or (\ref{skneg}) is being considered.  For further analysis of (\ref{vnhodZ}), we will also need to write 
the reciprocal of $Q$ in the form 
\beq\label{recipQ} 
\frac{1}{Q(Z)}=\frac{1}{2}\left(\frac{\wp'(W_+)}{\wp(Z)-\wp(W_+)}+ \frac{\wp'(W_-)}{\wp(Z)-\wp(W_-)} \right), 
\eeq
where from (\ref{QRrel}) and (\ref{Rtrav}) it follows that $1/Q(Z)$ has simple poles at values of $Z$ congruent to 
$\pm W_\pm \bmod \La$ (with $\La$ denoting the period lattice of the $\wp$ function), which are determined 
from the requirements 
\beq\label{Wpm}
\wp(W_+)+\wp(W_-)=2\wp(W), \qquad  \wp(W_+)\wp(W_-)=\wp(W)^2+\frac{1}{4\tilde{c}}. 
\eeq  
These two equations for $\wp(W_\pm)$ together imply that 
\beq\label{delsq}
\big(\wp(W_+)-\wp(W_-)\big)^2=-\frac{1}{\tilde{c}}.
\eeq 
Then from evaluating $\sfrac{\rd R}{\rd Z}$ at $Z=W_\pm$ and using (\ref{Rode}) 
together with the fact that $R^2+\tilde{c}=0$ at these points, we 
find that 
$$ 
\wp'(W_+)^2=\wp'(W_-)^2=-\frac{c^2}{4\tilde{c}^3}; 
$$ 
thus from (\ref{delsq}) we can fix signs so that 
\beq\label{Wpms}
\wp'(W_+)=-\wp'(W_-)=\frac{c}{2\tilde{c}^2}\big(\wp(W_+)-\wp(W_-)\big)^{-1},
\eeq
which ensures that $1/Q(Z)$ has residue $1/2$ at points congruent to $W_{\pm} \bmod\La$, and residue 
  $-1/2$ at points congruent to $-W_{\pm} \bmod\La$, as in the formula (\ref{recipQ}). 
\begin{thm}\label{vntrav} 
The smooth  travelling wave solutions (\ref{vntravwaves}) of Novikov's equation (\ref{novikov}) 
are given parametrically by $U=U(Z)$, $z=z(Z)$, where 
\beq\label{UQR} 
U(Z)=\frac{\pm\sqrt{c}\,R(Z)}{\sqrt{R(Z)^2+\tilde{c}}}
\eeq 
with $R(Z)$  defined by 
(\ref{Rtrav})  (up to the freedom to shift 
$Z\to Z+\,$const), and 
\beq\label{zvn}
z(Z)=\tfrac{1}{2}\log\left(\frac{\si(Z-W_+)\si(Z-W_-)}{\si(Z+W_+)\si(Z+W_-)}\right)+\big(\ze (W_+)+\ze (W_-) \big)Z+\mathrm{const},
\eeq 
with the parameters being specified by (\ref{vnparam}) and (\ref{vnc}), together with (\ref{Wpm}) and (\ref{Wpms}). 
\end{thm} 

\begin{prf}
The definition of $q$ and $r$ in (\ref{qr}) implies that $u^2=r^2/q$, so reducing to the travelling wave solutions 
and taking a square root produces 
$$U=\pm\frac{R}{\sqrt{Q}}, 
$$ 
where either choice of sign is valid. (The PDE (\ref{novikov}) is invariant under $u\to -u$.) Upon using 
(\ref{QRrel}) to replace $Q$ in terms of $R$, the expression (\ref{UQR}) results.  
As for the formula (\ref{zvn}), this follows from (\ref{vnhodZ}), using (\ref{recipQ}) and standard 
identities for Weierstrass functions to perform the 
integral $z=\int Q(Z)^{-1}\,\rd Z+\,$const.
\end{prf}

\begin{figure}
\centering 
\begin{subfigure}{.5\textwidth}
 \centering
 \includegraphics[width=.7\linewidth]{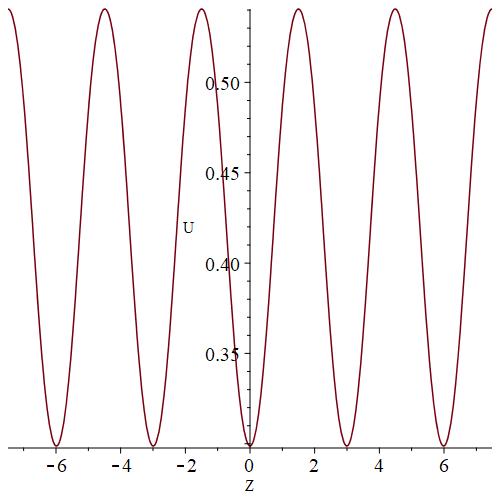}
 \caption{$U(Z)$ for $-5\om_1\leq Z\leq 5\om_1$.}\label{vnUvsZ} 
\end{subfigure}%
\begin{subfigure}{.5\textwidth}
 \centering
 \includegraphics[width=.7\linewidth]{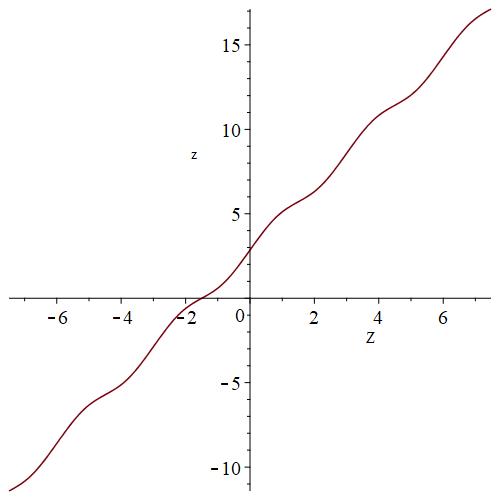}
 \caption{$z(Z)$ for $-5\om_1\leq Z\leq 5\om_1$.}\label{vnzvsZ} 
\end{subfigure}\caption{Parametric form of  travelling waves for Novikov's equation with real $Z$.}\label{vnUztravplot}
\end{figure}

\begin{exa}\label{vntravplot}
For illustration of the above  theorem, we use it 
to plot a particular travelling wave solution of (\ref{novikov}) which is bounded and real for $x,t\in\R$.
We choose the same Weierstrass cubic as in Example \ref{mchtravplot} by fixing the values of the invariants as before, but make a different choice of the 
parameter $W$, with an exact value of $\wp(W)$, namely    
$$ 
g_2=4, \qquad g_3=-1, \qquad \wp(W)=1, 
$$ 
which 
arises from taking $W\approx 1.134273216$. In this case, $\wp'(W)=-1$ and $\wp''(W)=4$, so 
from  (\ref{vnparam}) and  (\ref{vnc}) this gives the value of the velocity of the travelling wave and 
the other constants appearing in the solution as 
$$ 
c=\frac{\sqrt{23}}{8\sqrt{2}}, \qquad \tilde{c}=\frac{1}{8}, \qquad e=\frac{3}{4}.
$$ 
As before, we take three half-periods for the Weierstrass $\wp$ function given by 
$$ 
\om_1\approx 1.496729323 
, 
\qquad 
\om_2\approx 1.225694691 \ri 
, \qquad \om_3=\om_1+\om_2, 
$$
and avoid poles for real values of $Z$,  by exploiting the freedom to shift $Z$ and $z$ by constants in Theorem \ref{vntrav}, 
replacing $Z\to Z+\om_3$ in (\ref{Rtrav}) and (\ref{zvn}), and ensuring that $z$ is real for all 
$Z\in\R$ by an appropriate choice of constant in  (\ref{zvn}). 
Note that in this case we have 
$$ 
\wp(W_\pm)=1\pm \sqrt{2}\ri, 
$$ 
and we can choose $W_\pm$ to be complex conjugates of one another, so that 
$$ 
W_+=\overline{W}_- \approx 0.6575861671 - 0.3645241628\ri, 
$$ 
and $\wp'(W_+)=-\sqrt{23}\ri$ in accordance with (\ref{Wpms}). 
Then $U$ given by (\ref{UQR}) is a bounded periodic function for real argument $Z$, 
and the corresponding function $U(z)$ defined parametrically by $z(Z)$ is a bounded periodic travelling wave profile for  (\ref{novikov}). 
Indeed, using the quasiperiodicity 
of the Weierstrass sigma function, 
the formula (\ref{zvn}) shows that the period of $U(z)$ is 
$$
z(Z+2\om_1)-z(Z)=2\Big(\om_1\big(\ze(W_+)+\ze(W_-)\big)-(W_++W_-)\ze(\om_1)\Big)
\approx5.708708303$$ in this numerical example. 
\begin{figure}
 \centering
\label{Uztrw}
\epsfig{file=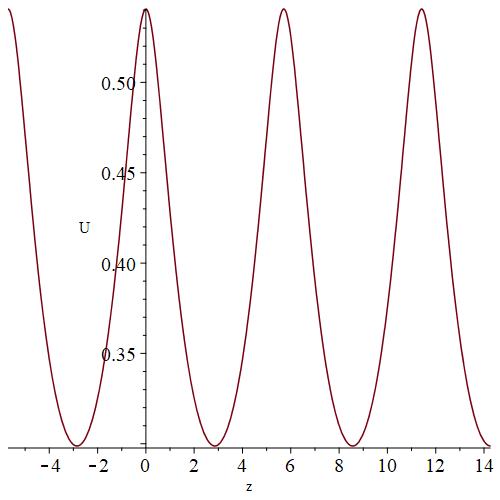, height=2.7in, width=2.7in}
\caption{Periodic travelling wave solution of Novikov's equation.
}
\end{figure}%
Clearly, in this case we have $Q(Z)=c^{-1}\big(R(Z)^2+\tilde{c}\big)>0$  for all real $Z$, so $\sfrac{\rd z}{\rd Z}=1/Q>0$ 
and $z(Z)$ is a monotone increasing function of its argument, as is visible from the right-hand panel of Fig.4. 
The left-hand panel of the latter figure shows $U$  plotted against $Z$, where both plots are for 
$-5\om_1\leq Z\leq 5\om_1$, while in Fig.5 
we have plotted $U$ against $z$, in the range $-3\om_1\leq Z\leq 4\om_1$, corresponding to the travelling wave profile for Novikov's equation. 
\end{exa}

\begin{remark}\label{dpvn} 
Thus far we have not discussed how the result of Theorem \ref{vntrav}, which describes  the travelling waves of 
Novikov's equation (\ref{novikov}), is related to the travelling waves of the Degasperis-Procesi equation (\ref{dp}). 
The connection between them is somewhat indirect, arising from the two reciprocal transformations (\ref{dprt}) 
and (\ref{vnrt}), together with the two Miura maps (\ref{kkskmiura}), and their reductions to the travelling wave solutions. 
On the one hand, 
as shown in \cite{bh}, for the travelling wave solution (\ref{negkktrav}) of (\ref{dp3rd}), 
or equivalently (\ref{VTdp}), there is a corresponding  
Kaup-Kupershmidt field $V=V(Z)=\sfrac{3}{4}d^{-1}P(Z)^2+\,$const. From the formulae (\ref{alphatrav}) together with (\ref{negkktrav}), 
it is apparent that $V$ is an elliptic function of $Z$ with double poles only at points congruent to $\pm W_2$, with the leading order in the 
Laurent expansion at these points  being $V(Z)=-\sfrac{3}{4}(Z\mp W_2)^{-2}+O(1)$, so fixing the value at $Z=0$ we find that it can be 
written in the form 
\beq\label{kk1} 
V(Z)=-\sfrac{3}{4}\wp(Z+W_2)-\sfrac{3}{4}\wp(Z-W_2)-\sfrac{3}{2}\wp(W_2)+\sfrac{3}{4}\al^2\wp'(W_2)^2
\eeq
(but see equation (2.21) in \cite{bh} for another equivalent expression).
On the other hand, from the formula  (\ref{recipQ}) we can use (\ref{vqform}) to produce a (reduced) modified 
field variable $v=v(Z)$ given by 
\beq\label{skvred} 
\begin{array}{rcl}
v(Z) &= &\frac{1}{2}\frac{\rd}{\rd Z}\log Q(Z)^{-1}+\frac{1}{Q(Z)} \\ 
& =& -\frac{1}{2}\left(\frac{\wp'(Z)-\wp'(W_+)}{\wp(Z)-\wp(W_+)}+\frac{\wp'(Z)-\wp'(W_-)}{\wp(Z)-\wp(W_-)}\right) \\
& = & 2\ze(Z)-\ze(Z+W_+)+\ze(W_+)-\ze(Z+W_-)+\ze(W_-).
\end{array}
\eeq 
(The latter form of $v(Z)$ arises from choosing the second term in (\ref{vqform}) with a  plus sign, and applying suitable elliptic function 
identities; choosing the minus sign instead just replaces $W_\pm\to-W_\pm$ in the above expression.) So to the solution (\ref{Rtrav}) of the ODE 
(\ref{Rode}) for travelling waves of the system  (\ref{skneg}) there corresponds a Sawada-Kotera field $\hat{V}=\hat{V}(Z)$ 
given by 
\beq\label{sk2} 
\hat{V}(Z)=\frac{\rd v}{\rd Z}-v^2=-6\wp(Z)-6\wp(W), 
\eeq 
where the preceding explicit formula is obtained by considering the leading terms in the Laurent expansions of $v(Z)$ around its simple poles
at points congruent to $Z=0,-W_+,-W_-\bmod \La$, where it has residues $2,-1,-1$ respectively; then noting that consequently $\hat{V}(Z)$ 
has only double poles at points congruent to $Z=0$, with leading order $\hat{V}(Z)\sim-6Z^{-2}-3\big(\wp(W_+)+\wp(W_-)\big)$, 
and using (\ref{Wpm}), the final expression in (\ref{sk2}) follows. Now by reducing the first Miura map in  (\ref{kkskmiura}) to these travelling waves, 
there should be an associated Kaup-Kaupershmidt field $V(Z)$, which is found from 
$$ 
V(Z) 
=-\tfrac{1}{2}\frac{\rd v}{\rd Z}-\tfrac{1}{4}v^2 
=\tfrac{1}{4}\hat{V}(Z)-\tfrac{3}{4}\frac{\rd v}{\rd Z}, 
$$
so inserting the expressions  (\ref{skvred}) and (\ref{sk2})  this yields 
\beq\label{kk2}
V(Z)  =
-\sfrac{3}{4}\wp(Z+W_+)-\sfrac{3}{4}\wp(Z+W_-)-\sfrac{3}{2}\wp(W).
\eeq 
The connection between the Degasperis-Procesi/Novikov travelling wave solutions is now established by showing that it is 
consistent to identify the two different formulae (\ref{kk1}) and (\ref{kk2}) for a Kaup-Kupershmidt field $V$. First of all, 
the precise form of these travelling waves is specified up to the freedom to shift the independent variable $Z$ by an arbitrary constant, 
so if we replace $Z\to Z-\sfrac{1}{2}(W_++W_-)$ in (\ref{kk2}), then the double poles in the solution are at points 
congruent to $\pm \sfrac{1}{2}(W_+-W_-)\in\C /\La$. Hence, comparing with (\ref{kk1}), we can identify
\beq\label{W2}
W_2=\tfrac{1}{2}(W_+-W_-).
\eeq 
As a consequence of the duplication formula for the $\wp$ function, doubling (\ref{W2}) gives 
$$ 
\wp(W_+-W_-)=\wp(2W_2)=-2\wp(W_2)+\tfrac{1}{4}\left(\frac{\wp''(W_2)}{\wp'(W_2)}\right)^2, 
$$ 
while at the same time,  the addition formula for $\wp$ together with  (\ref{Wpms}) implies that 
$$ 
\wp(W_+-W_-)=-\wp(W_+)-\wp(W_-).
$$ 
So combining the latter two results with  (\ref{Wpm}) and the first expression for $\al$ in (\ref{alphatrav}), we obtain the equality 
\beq\label{PW} 
\wp(W)=\wp(W_2)-\tfrac{1}{2}\al^2\wp'(W_2)^2,
\eeq 
which implies that we can identify the constant terms in the  two formulae (\ref{kk1}) and (\ref{kk2}). Finally, for these two different 
expressions for  $V$ to be compatible, we require that the independent variable $Z$  should be the same in each case: in  
(\ref{kk1}) it is given by $X-dT$, while in (\ref{kk2}) it is $X-\tilde{c}T$, so (assuming that $X,T$ are the same in both cases) 
this means 
that the two wave  velocities should coincide. Then from (\ref{vnparam}) we may make use of (\ref{PW}) to write 
$$
\tilde{c}^{-1}=2\wp''(W)=12\wp(W)^2-g_2=12\big(\wp(W_2)-\tfrac{1}{2}\al^2\wp'(W_2)^2\big)^2-g_2, 
$$ 
so that 
\beq\label{tcinv}
\tilde{c}^{-1}=2\wp''(W_2)-12\al^2\wp(W_2)\wp'(W_2)^2+3\al^4\wp'(W_2)^4, 
\eeq 
whereas from (\ref{negkktrav}) and (\ref{alphatrav}) we have 
\beq\label{dinv}
\begin{array}{rcl}
d^{-1}& =& -\al^4\wp'(W_2)^2\wp'(W_1)^2 
\\ 
&=&-\al^4\wp'(W_2)^2\Big(\wp'(W_2)^2+4\big(\wp(W_1)^3-\wp(W_2)^3\big)-g_2\big(\wp(W_1)-\wp(W_2)\big)\Big), 
\end{array} 
\eeq 
using the first order ODE for the $\wp$ function. Finally, the second expression for $\al$ in (\ref{alphatrav}) allows us to substitute 
$\wp(W_1)=\wp(W_2)+\al^{-1}$ in (\ref{dinv}), and then eliminate $g_2=12\wp(W_2)^2-2\wp''(W_2)$, followed by replacing 
$\wp''(W_2)=-2\al\wp'(W_2)^2$ in the resulting formula for $d^{-1}$ and doing the same in (\ref{tcinv}), which results in 
$$ 
d^{-1}=-4\al\wp'(W_2)^2-12\al^2\wp(W_2)\wp'(W_2)^2+3\al^4\wp'(W_2)^4
=\tilde{c}^{-1}; 
$$
so the wave velocities are the same, 
as required. 

We have already briefly commented on the discrete symmetry associated with the choice of sign in (\ref{vqform}), 
which at the level of the travelling wave reduction of (\ref{skneg}) produces two different modified 
variables 
\beq\label{vsym} 
v_\pm = -\sfrac{1}{2}(\log Q)_Z\pm Q^{-1}, 
\eeq where $v_+$ is given by (\ref{skvred}), and $v_-$ is given by the same formula but with 
$W_+\to-W_+$, $W_-\to -W_-$ throughout. 
The Miura map $\sfrac{\rd v_\pm}{\rd Z}-v_\pm^2$ 
gives the same (reduced) Sawada-Kotera field $\hat{V}(Z)$, as can be observed directly from 
the elliptic function expression on the far right-hand side of (\ref{sk2}):  
this is invariant under changing the signs of $W_\pm$. However, applying the other Miura map to $v_\pm$ produces two different
reduced Kaup-Kupershmidt fields, namely 
\beq\label{Vplusminus}
V_\pm (Z)= -\sfrac{1}{2}\tfrac{\rd v_\pm}{\rd Z}-\sfrac{1}{4}v_\pm^2=-\sfrac{3}{4}\wp(Z\pm W_+)-\sfrac{3}{4}\wp(Z\pm W_-)-\sfrac{3}{2}\wp(W)
\eeq  
(with (\ref{kk2}) just being the first of these).
So far, in order to derive the results on parametric travelling waves in 
Theorem \ref{vntrav}, we have not needed to make use of this discrete symmetry, but in the case of the scaling similarity solutions 
considered in the next subsection 
it will be a more vital ingredient. 
\end{remark}

\subsection{Scaling similarity solutions} 

Novikov's equation (\ref{novikov}) 
has a one-parameter family of similarity solutions, for which both $u$ and the momentum density $m$  in (\ref{vn}) scale the same way, given by 
the same form of reduction as in the case of the 
mCH equation (\ref{mch}), that is   
\beq\label{vnsimred}
u(x,t)=t^{-\sfrac{1}{2}}U(z), \qquad m(x,t)=t^{-\sfrac{1}{2}}M(z), \qquad z=x+\al\log t. 
\eeq 
This reduction results in an autonomous ODE of third order for $U(z)$, namely 
\beq\label{vnU3rd} 
(U^2+\al)(U_{zzz}-U_z)+(3UU_z-\sfrac{1}{2})(U_{zz}-U)=0.
\eeq 
To obtain solutions of the latter ODE in parametric form, we will 
consider corresponding similarity solutions of the negative Sawada-Kotera flow  
(\ref{skneg}), related via the reciprocal transformation (\ref{vnrt}). 
Under the reduction (\ref{vnsimred}), the quantities $q,r$ given by (\ref{qr}), that appear 
in the associated conservation law 
(\ref{vncons}), take the form 
\beq\label{qrsimred} 
q(x,t)=t^{-\sfrac{1}{3}}Q(z), \qquad r(x,t)=t^{-\sfrac{2}{3}}R(z).
\eeq   

The system (\ref{skneg}) has scaling similarity solutions given by taking 
\beq\label{QRsimred}
q(X,T)= T^{-\sfrac{1}{3}}Q(Z), \qquad r(X,T)=T^{-\sfrac{2}{3}}R(Z), \qquad Z=XT^{\sfrac{1}{3}}.
\eeq 
Applying this similarity reduction means that the first equation in the system produces 
\beq\label{QRdZ} 
\frac{1}{3} \frac{\rd}{\rd Z} \big( ZQ^{-1}\big)= \frac{\rd}{\rd Z} \big(R^2Q^{-1}\big), 
\eeq 
while, using the definition of $\hat{V}$ in (\ref{Vhatdef}), the second equation becomes 
\beq\label{VhatQR} 
\frac{(Q^{1/2})_{ZZ}}{Q^{1/2}}+\frac{1}{Q^2}=\frac{R_{ZZ}+1}{R}. 
\eeq 
The equation (\ref{QRdZ}) integrates to give 
$$ 
Q^{-1}\big(\sfrac{1}{3}Z-R^2\big)=\mathrm{const}.
$$ 
If we 
let $\al$ denote the integration constant above, then 
this gives 
\beq\label{QRal} 
\al Q=\frac{Z}{3}-R^2, 
\eeq 
and we find that this corresponds precisely to the image under  the reciprocal transformation   
(\ref{vnrt}) of the similarity solutions (\ref{vnsimred}) of Novikov's equation (\ref{vn}), where 
(without loss of generality) we can set $t=T$ and perform a calculation analogous 
to (\ref{Zcalc}) to find the hodograph transformation 
\beq\label{Zscahod} 
\rd Z=Q\,\rd z,
\eeq 
so that the system consisting of (\ref{QRal}) and (\ref{VhatQR}) is a consequence 
of replacing the $z$ derivatives in  (\ref{vnU3rd}) with 
$\sfrac{\rd }{\rd z}=Q\sfrac{\rd }{\rd Z}$ and rewriting suitable combinations 
of $U$ and its derivatives in terms of the quantities $Q$ and $R$.

\begin{remark}\label{simpot} 
The form of the scaling similarity reduction (\ref{QRsimred}) and the equation (\ref{QRal}) arise from (\ref{qrphi}) by choosing a potential of the 
form 
$$ 
\Phi(X,T)=\varphi(Z)-\al\log T, \qquad Z=XT^{\sfrac{1}{3}}, 
$$ 
with $1/Q(Z)=\sfrac{\rd}{\rd Z}\varphi(Z)$, so that integrating 
(\ref{Zscahod}) gives $z=\varphi(Z)$, 
but in due course we will obtain a slightly more explicit formula for the potential $\varphi$ in terms of tau functions. 
\end{remark}

Guided by the results on travelling waves in the preceding subsection, we next use (\ref{QRal}) 
with $\al\neq 0$ 
to substitute for $Q$ and $Q_Z=\al^{-1}(\sfrac{1}{3}-2RR_Z)$, 
$Q_{ZZ}=-2\al^{-1}(RR_{ZZ}+R_Z^2)$ in (\ref{VhatQR}), to find a single ODE 
of second order for $R(Z)$, that is 
\beq\label{RPV}
\frac{\rd^2 R}{\rd Z^2}=\frac{1}{R^2-\sfrac{1}{3}Z}\left(R\left(\frac{\rd R}{\rd Z}\right)^2-\frac{R^2}{Z}\left(\frac{\rd R}{\rd Z}\right)+\frac{3}{Z} R^4-2R^2+\frac{(\sfrac{1}{12}-3\al^2)}{Z}R+\sfrac{1}{3}Z\right) .
\eeq 
The above equation is very similar in form to the Painlev\'{e} V equation (\ref{pV}), 
and indeed it is related to it by a simple change of dependent and independent variables. 

\begin{lem}\label{Rpv}
The solutions $R=R(Z)$ of the ODE (\ref{RPV}) are given by 
\beq\label{Zwtrans} 
R=\sqrt{\frac{Z}{3}}\left(\frac{1+w}{1-w}\right), 
\eeq 
where $w=w(\ze)$ is a solution of the Painlev\'e V equation (\ref{pV}) with parameters 
\beq\label{pvco}
\tilde{\al}=\sfrac{1}{2}\al^2, 
\quad 
 \tilde{\be}=-\sfrac{1}{2}\al^2, 
\quad 
\tilde{\gam}=1, \quad 
\tilde{\delta}=0, 
\eeq 
and \beq\label{Zze} 
\ze=(4Z/3)^{\sfrac{3}{2}}
.
\eeq 
\end{lem}
\begin{prf}
To see this, note that the coefficient of $\big(\sfrac{\rd R}{\rd Z}\big)^2$ in 
(\ref{RPV}) is a rational function of $R$ of degree 2, with poles at $R=\pm\sqrt{\sfrac{Z}{3}}$, so 
the ODE has fixed singularities at these points and 
at $R=\infty$,
while the coefficient 
of   $\big(\sfrac{\rd w}{\rd \ze}\big)^2$ in 
(\ref{pV}) is 
\beq\label{wzesq}
\frac{1}{2w}+\frac{1}{w-1}=\frac{3w-1}{2w(w-1)}, 
\eeq 
which suggests transforming the dependent variable with the M\"{o}bius transformation
(\ref{Zwtrans}).
in order to move the fixed singularities to $w=0,1,\infty$. This transformation 
indeed produces the correct coefficient (\ref{wzesq}), transforming (\ref{RPV}) to an equation 
with leading terms 
$$ 
\frac{\rd^2 w}{\rd Z^2} = \left( \frac{1}{2w}+\frac{1}{w-1}\right)\left(\frac{\rd w}{\rd Z}\right)^2+\cdots, 
$$
and then  to obtain the precise form of Painlev\'{e} V further requires a  change of independent variables, namely 
the replacement 
$$
Z=\frac{3}{4}\ze^{\sfrac{2}{3}}, 
$$ 
with inverse (\ref{Zze}), which produces the equation (\ref{pV}) with the particular choice of coefficients 
(\ref{pvco}).
\end{prf}

It is a result due to Gromak that Painlev\'{e} V with $\tilde{\delta}=0$ can be solved in terms of Painlev\'e III 
transcendents \cite{gromak} (see also
$\S32.7$(vi) in \cite{dlmf}). If we replace the set of parameters in (\ref{piii}) with $(\hat{\al}, \hat{\be}, \hat{\gam}, \hat{\delta})$, 
and denote the dependent and independent variables by $\hat{w},\eta$, respectively, then a more exact statement is that, 
if $\hat{w}=\hat{w}(\eta)$ is a solution of   Painlev\'e III with parameters $(\hat{\al}, \hat{\be}, 1, -1)$, then $w=w(\ze)$ 
with 
\beq\label{p3p5} 
w={\cal F}\Big(\hat{w},\tfrac{\rd \hat{w}}{\rd \eta},\eta,\al\Big), \qquad \eta=\sqrt{2\ze}, 
\eeq
satisfies Painlev\'{e} V with parameters given by 
\beq\label{p5param}
(\tilde{\al}, \tilde{\be}, \tilde{\gam}, \tilde{\delta})=\Big(\sfrac{1}{32}(\hat{\be}-\varepsilon \hat{\al}+2)^2, -\sfrac{1}{32}(\hat{\be}+\varepsilon \hat{\al}-2)^2,-\varepsilon ,0\Big),
\eeq 
where $\varepsilon=\pm 1$ and ${\cal F}$ is a certain rational function of its arguments (see $\S32.7$(vi) in \cite{dlmf} for full details). We now wish 
to use Gromak's result in order 
to show that the solutions of the ODE (\ref{RPV}) are related to the scaling similarity solutions of the negative 
Kaup-Kupershmidt flow (\ref{VTdp}), 
which in turn correspond to solutions of the Degasperis-Procesi equation (\ref{dp}) via the reciprocal transformation (\ref{dprt}). 
In \cite{bh} it was explained how the scaling  similarity reduction of (\ref{VTdp}), or rather (\ref{dp3rd}), 
results in an ODE which is equivalent to  Painlev\'e III with parameter values 
\beq\label{p3redpara}
(\hat{\al}, \hat{\be}, \hat{\gam}, \hat{\delta})=(0, \sfrac{4}{3}a, 1, -1), 
\eeq 
where $a$ is arbitrary. However, applying the transformation (\ref{p3p5}) directly to the latter solutions with $\hat{\al}=0$ does not lead to 
solutions of Painlev\'{e} V with $\tilde{\al}=-\tilde{\be}$, which is what we require from (\ref{pvco}). 
Thus, in order to obtain the required connection, we can apply one of the B\"acklund transformations 
for Painlev\'e III (see e.g. \cite{bcm}), which 
sends 
\beq\label{p3shift} 
(\hat{\al}, \hat{\be}, 1, -1)\longrightarrow 
(\hat{\al}-2, \hat{\be}+2, 1, -1) .
\eeq 
Then starting from the parameter values (\ref{p3redpara}) and applying (\ref{p3shift}) followed 
by the transformation (\ref{p3p5}) in the case $\varepsilon=-1$, a solution of  Painlev\'{e} V with the 
appropriate parameters arises. 

For the similarity reductions of Novikov's equation, it is more convenient to describe 
these connections directly in terms of the solutions of the ODE 
\beq\label{dp3rdred} 
 \frac{\rd^2 P}{\rd Z^2}=\frac{1}{P}\left(\frac{\rd P}{\rd Z}\right)^2-\frac{1}{Z}\left(\frac{\rd P}{\rd Z}\right)+\frac{1}{Z}\big( 3P^3+a) -\frac{1}{P}, 
\eeq   
which was derived in \cite{bh} by taking scaling similarity solutions of (\ref{dp3rd}), of the form 
\beq\label{dpred} 
p(X,T)=T^{-1/3}P(Z), \qquad Z=XT^{1/3}. 
\eeq 

\begin{propn}\label{PRprop}
Each solution $P=P(Z)$ of the ODE (\ref{dp3rdred}) with parameter 
\beq\label{aalpha} 
a=-\tfrac{3}{2}\pm 3\al
\eeq 
provides a solution $R=R(Z)$ of (\ref{RPV}) with parameter $\al$, via the formula 
\beq\label{RP}  
R=\frac{Z(P_Z+1)}{3P^2}-\frac{(a+1)}{3P}, 
\eeq 
and conversely, each solution of (\ref{RPV}) provides a solution of  (\ref{dp3rdred})
with parameter $a$ given by (\ref{aalpha}), according to the formula 
\beq\label{PR}
P=-\frac{ZR_Z+(\pm 3\al -\sfrac{1}{2})R}{3R^2-Z}. 
\eeq 
\end{propn}
\begin{prf}
The ODE (\ref{dp3rdred}) corresponds to  Painlev\'e III with parameter values 
(\ref{p3redpara}), via the transformation 
\beq\label{hatwP} 
\hat{w}(\eta)=(Z/3)^{-\sfrac{1}{4}}P(Z), \qquad \eta = 4(Z/3)^{\sfrac{3}{4}}, 
\eeq  
as given (with  slightly different notation) in equation (3.21) in \cite{bh}. 
Starting from a solution $\hat{w}$ of Painlev\'e III with these values of 
parameters, the shift (\ref{p3shift}) is achieved by the transformation 
\beq\label{wwstar}
w^*=-\frac{1}{\hat{w}}+\frac{\hat{\be}+2}{\eta(\hat{w}_\eta-\hat{w}^2+1)+\hat{w}}
\eeq 
(cf.\ \cite{bcm} and equation (3.23) in \cite{bh}), 
producing a new solution $w^*(\eta)$ for parameter values $(-2,\hat{\be}+2,1,-1)$. 
Then the corresponding solution of  Painlev\'{e} V 
is obtained by applying the transformation (\ref{p3p5}) to $w^*$, which gives 
\beq\label{wvstar} 
w={\cal F}\Big(w^*,\tfrac{\rd w^*}{\rd \eta},\eta,-2\Big)=\frac{v^*-1}{v^*+1}, 
\eeq 
where (from $\S32.7$(vi) in \cite{dlmf}) 
\beq\label{vstar} 
v^*=w^*_\eta+(w^*)^2-\eta^{-1}w^*, 
\eeq 
so that $w=w(\ze)$ satisfies (\ref{pV}) with parameters 
\beq\label{p5coeffs} 
(\tilde{\al}, \tilde{\be}, \tilde{\gam}, \tilde{\delta})=\Big(\sfrac{1}{32}(\be+2)^2, -\sfrac{1}{32}(\be+2)^2,1,0\Big),
\eeq  
as found by replacing $\hat{\al}\to -2$, $\hat{\be}\to\hat{\be+2}$ and $\varepsilon\to -1$ in  (\ref{p5param}). 
The transformation rule $ \eta=\sqrt{2\ze}$ for the independent variables is consistent with the expressions for 
$\zeta,\eta$ in terms of $Z$, as presented in (\ref{Zze}) and (\ref{hatwP}), respectively. Thus we can 
rewrite the expression on the far right-hand side of (\ref{wvstar}) in terms of $\hat{w}$ and its first derivative $\hat{w}_\eta$, by using the B\"acklund 
transformation 
(\ref{wwstar}) together with the ODE (\ref{piii}) for $\hat{w}$, to eliminate the second derivative, and then use the formulae 
(\ref{Zwtrans}) and (\ref{hatwP}) to write the left- and right-hand sides in terms of $Z$ and $R,P$ 
respectively. An immediate simplification can be made by noting that the M\"obius transformation of 
$w$ in   (\ref{Zwtrans})  is just the inverse of the M\"obius transformation of  $v^*$ in (\ref{wvstar}), 
which implies that $R=\sqrt{\sfrac{Z}{3}}\,v^*$, so it is only necessary to rewrite $v^*$ given by  (\ref{vstar}) 
as a rational function of $\hat{w}$ and $\hat{w}_\eta$, before applying  (\ref{hatwP}) to obtain (\ref{RP}). 
The relationship between the parameters in  (\ref{dp3rdred}) and (\ref{RPV}) arises by comparing  
(\ref{p5coeffs}) with (\ref{pvco}), which gives $(\hat{\be}+2)^2=16\al^2$; then 
setting $\hat{\be}=\sfrac{4}{3}a$ from (\ref{p3redpara}) and taking a square root, (\ref{aalpha}) follows. 
For the converse, one can differentiate both sides (\ref{RP}) with respect to $Z$ and use  (\ref{dp3rdred}) 
to eliminate the $P_{ZZ}$ term, which produces a pair of equations for $R$ and $R_Z$ as rational functions 
of $P$ and $P_Z$. After eliminating $P_Z$ from these two equations, the expression (\ref{PR}) for 
$P$ in terms of $R$ and $R_Z$ results by replacing $a$ from (\ref{aalpha}), with either choice of sign. 
\end{prf}

In addition to the shift (\ref{p3shift}), Painlev\'e III with $\hat{\gam}=-\hat{\delta}=1$ admits another elementary 
B\"acklund 
transformation which sends $\hat{\al}\to\hat{\al}+2$, $\hat{\be}\to\hat{\be}+2$ \cite{bcm}. 
We remarked in \cite{bh} that it is necessary to take the composition of 
these two transformations, sending $\hat{\be}\to\hat{\be}+4$ and leaving $\hat{\al}$ fixed, 
in order to preserve the condition $\hat{\al}=0$ required 
for the  parameter values 
(\ref{p3redpara}) associated  with (\ref{dp3rdred}). Furthermore, in \cite{bh} (see Table 2 therein) we also 
applied this composition of two Painlev\'e III B\"acklund 
transformations, which has the effect of sending $a\to a+3$,  to generate the first few members of a sequence of algebraic solutions 
of  (\ref{dp3rdred}) for parameter values $a=3n$, $n\in\Z$. Here we now show how Proposition \ref{PRprop} 
leads to a direct derivation of the corresponding B\"acklund 
transformation for (\ref{dp3rdred}). To present these results, it will be convenient to denote a solution 
of  (\ref{dp3rdred}) with parameter value $a$ by $P_a=P_a(Z)$. 

\begin{cor}\label{PodeBTs} The equation  (\ref{dp3rdred}) admits two elementary B\"acklund 
transformations, 
given by 
\beq\label{signa} 
P_{-a}=-P_a 
\eeq
and 
\beq\label{shifta} 
P_{a+3}=-P_a-\frac{(2a+3)R_a}{3R_a^2-Z},  
\eeq
where 
\beq\label{Ra}
R_a=
\frac{Z\left(\tfrac{\rd P_a}{\rd Z}+1\right)}{3P_a^2}-\frac{(a+1)}{3P_a}. 
\eeq 
\end{cor} 
\begin{prf} The first transformation (\ref{signa}) is an immediate consequence of 
the invariance of the ODE under $P\to -P$, $a\to -a$. As for the second one, 
note that there is an arbitrary choice of sign in (\ref{aalpha}), because (\ref{RPV}) depends 
only on the square of $\al$, and 
without loss of generality we can fix
\beq\label{alfix} 
\al = \sfrac{1}{3}a+\sfrac{1}{2}
\eeq 
Then from the fact that the 
symmetry $a\to-a-3$ sends $\al\to -\al$, we see that 
\beq\label{Rsame} 
R_{-a-3}=R_a, 
\eeq
or in other words, there are two solutions of  (\ref{dp3rdred}) that produce the 
same solution of (\ref{RPV}), namely (for the same $R$) we have $P_a$ given by taking  the plus sign 
in (\ref{PR}), and $P_{-a-3}$ given by  taking  the minus sign. If we subtract these two expressions then we obtain 
$$ 
P_a-P_{-a-3}=-\frac{6\al R}{3R^2-Z} \qquad \mathrm{with} \,\,\, R=R_a,
$$ 
and then applying the transformation (\ref{signa}) to the second term on the left-hand side and 
substituting for $\al$ with (\ref{alfix}), the result (\ref{shifta}) follows.
\end{prf}

We now explain how the discrete symmetry (\ref{Rsame}) of the ODE (\ref{RPV}), given by 
sending $a\to-a-3$, or equivalently $\al\to -\al$, corresponds to 
changing the sign  of the second term in (\ref{vqform}), at the level of the scaling similarity 
solutions of (\ref{skneg}). From applying the reduction (\ref{QRsimred}) to the latter system, we can remove a 
factor of $T^{1/3}$ to obtain reduced  modified variables $v_\pm=v_\pm(Z)$ 
that are expressed in terms of $Q$ 
by the same 
formula (\ref{vsym}) as in the travelling wave case. Then on the one hand, 
(by an abuse of notation) we can replace the Sawada-Kotera field $\hat{V}(X,T)\to T^{2/3}\hat{V}(Z)$, 
where the reduced field is given by 
the Miura formula 
\beq \label{Vhatsk}
\hat{V}(Z)=\frac{\rd v_+}{\rd Z}-v_+^2 = \frac{\rd v_-}{\rd Z}-v_-^2;
\eeq 
while on the other hand, applying the same scaling $V(X,T)\to T^{2/3}V(Z)$ 
to the Kaup-Kupershmidt field, the other Miura map 
gives two different reduced fields, namely  
\beq\label{Vpmkk} 
V_\pm (Z) =-\tfrac{1}{2}\frac{\rd v_\pm}{\rd Z}-\tfrac{1}{4}v_\pm^2.
\eeq
From (\ref{vsym}), 
the above formula defines each of $V_\pm$ as a rational function of $Q$ and its derivatives, which in turn 
can be written as a rational function  of $R$ and its first derivative, by using (\ref{QRal}) to substitute 
$Q=(\sfrac{1}{3}Z-R^2)/\al$, and using (\ref{RPV}) to eliminate the second derivative of $R$; 
the resulting  expression is somewhat unwieldy and is omitted here. (Some of these calculations 
are best verified with computer algebra.) However, a further calculation, using 
(\ref{RP}) to replace $R$ and $R_Z$ in terms of $P$ and $P_Z$ and $\al$ by (\ref{alfix}), 
with (\ref{dp3rdred}) used to replace $P_{ZZ}$ terms, produces the much more compact formula 
\beq\label{Vkk}
V_+=-\frac{1}{4P^2}\left(\Big(\frac{\rd P}{\rd Z}\Big)^2-1\right) 
+ \frac{1}{2ZP}\left( 
\frac{\rd P}{\rd Z}-3P^3-a 
\right) 
,
\eeq
where $P=P_a$ above. 

The right-hand side of the expression (\ref{Vkk}) for the (reduced) Kaup-Kupershmidt field $V_+$ coincides with the case $b=3$ of equation (3.12) in \cite{bh}, where it was 
derived by applying the scaling similarity reduction (\ref{dpred}) to the equation (\ref{ep}) - recall that this same relation defines $V$ in terms of $p$ in both the negative KdV 
and  Kaup-Kupershmidt flows.   Similarly, the same calculation for $V_-$ begins by replacing every occurrence of 
$\al$ by $-\al$, and results in the same expression as (\ref{Vkk}) but with $P\to P_{-a-3}$, $a\to -a-3$. Thus we can write 
$$ 
V_+=V_a, \qquad V_-=V_{-a-3}, 
$$ 
where 
\beq\label{Vkka}
V_a=-\frac{1}{4P_a^2}\left(\Big(\frac{\rd P_a}{\rd Z}\Big)^2-1\right) 
+ \frac{1}{2ZP_a}\left( 
\frac{\rd P_a}{\rd Z}-3P_a^3-a 
\right) 
.
\eeq
Analogously, for a fixed value of the  parameter $a$ we can also express the Sawada-Kotera field $\hat{V}$ in terms of $P=P_a$, and denote the result by $\hat{V}_a$, 
that is 
$$ 
\hat{V}_a= -\frac{1}{P_a^2}\left(\frac{\rd P_a}{\rd Z}+2\right) \left(\frac{\rd P_a}{\rd Z}+1\right) +\frac{1}{ZP_a}\left(-\frac{\rd P_a}{\rd Z}+3P_a^3+a\right) 
; 
$$
but then due 
to the equality of the two different Miura expressions in (\ref{Vhatsk}), we have that 
\beq\label{Vsyms}
\hat{V}_a=\hat{V}_{-a-3}, \qquad \mathrm{and}\qquad V_a=V_{-a}, 
\eeq 
where the latter identity follows from the invariance of (\ref{Vkka}) under $P_a\to P_{-a}= -P_a$, $a\to -a$.

For what follows, we also need to introduce tau functions $\tau_a(Z),\hat{\tau}_a(Z)$ associated with the reduced Kaup-Kupershmidt/Sawada-Kotera fields, respectively, 
which are defined by 
\beq\label{taukksk} 
V_a(Z)=\tfrac{3}{4}\frac{\rd^2}{\rd Z^2}\log \tau_a(Z), \qquad 
\hat{V}_a(Z)=6\frac{\rd^2}{\rd Z^2}\log \hat{\tau}_a(Z).
\eeq 
The above 
definition implies that $\tau_a(Z)$ has a movable simple zero at any point  $Z=Z_0\neq 0$ where $V_a(Z)$ has a movable 
double pole (with the local Laurent expansion being $V_a(Z)=-\sfrac{3}{4}(Z-Z_0)^{-2}+O(1)$ there), and an analogous relationship holds between  $\hat{\tau}_a(Z)$
and $\hat{V}(Z)$. From the above definition, together with the identity 
\beq\label{VVhat}
V_{\pm}=\tfrac{1}{4}\hat{V}-\tfrac{3}{4}\frac{\rd v_{\pm}}{\rd Z}
\eeq 
(which we made use of before, as part of the discussion of travelling waves  in Remark \ref{dpvn}), 
for a suitable choice of gauge we can also express the two modified fields $v_\pm$ in terms of these tau functions
as 
\beq\label{vpmtau}
v_+(Z)=\frac{\rd}{\rd Z}\log \left(\frac{\hat{\tau}_a(Z)^2}{\tau_a(Z)}\right), \qquad 
v_-(Z)=\frac{\rd}{\rd Z}\log \left(\frac{\hat{\tau}_{-a-3}(Z)^2}{\tau_{-a-3}(Z)}\right),
\eeq 
and from (\ref{Vsyms}) we can identify 
\beq\label{tauids}
\hat{\tau}_a(Z)=\hat{\tau}_{-a-3}(Z), \qquad 
\tau_a(Z)=\tau_{-a}(Z). 
\eeq 
All the ingredients required to state the main result on scaling similarity solutions 
are now in place.

\begin{figure}
 \centering
\label{Uzvnfig}
\epsfig{file=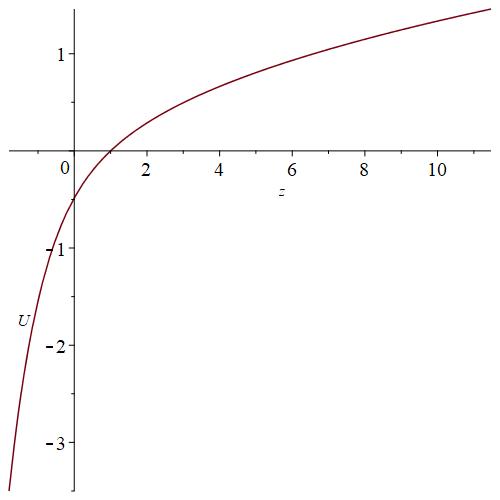, height=2.7in, width=2.7in}
\caption{Plot of $U$ against $z$ for the parametric solution 
(\ref{Uzparamvn}) with $0.505\leq \eta\leq 10$.
}
\end{figure} 

\begin{thm}\label{vnscalparam} 
The solutions of the ODE (\ref{vnU3rd}) for the similarity reduction 
(\ref{vnsimred}) of Novikov's equation (\ref{novikov}), with $\al\neq 0$, 
are given parametrically by $U=U(Z)$, $z=z(Z)$,  where $U$ is defined by 
\beq\label{UQRPV} 
U(Z)=\frac{\pm\sqrt{\al}\,R(Z)}{\sqrt{\sfrac{Z}{3}-R(Z)^2}}
,
\eeq 
with $R(Z)$  being a solution of the ODE 
(\ref{RPV}),  and 
\beq\label{zvntaus}
z(Z)=\tfrac{1}{2}\log \tau_{a+3}(Z) -\tfrac{1}{2}\log \tau_{a}(Z)+\mathrm{const},
\eeq 
in terms of two reduced Kaup-Kupershmidt tau functions $\tau_a$, $\tau_{a+3}$ connected via the B\"acklund transformation 
(\ref{shifta}) for (\ref{dp3rdred}), 
and $a=3\big(\al-\sfrac{1}{2}\big)$. 
\end{thm} 
\begin{prf} By (\ref{qr}), (\ref{vnsimred}) 
and (\ref{qrsimred}), we have $U(z)=\pm R(z)/\sqrt{Q(z)}$, so to give the solutions in parametric form we consider $z=z(Z)$ and 
(by the  usual abuse of notation) denote the associated functions 
with argument $Z$ by the same letters, so that (\ref{UQRPV}) follows directly from  (\ref{QRal}) after taking a square root. 
Then from combining  (\ref{vsym}) and (\ref{vpmtau}) we find 
$$ 
\tfrac{1}{2}(v_+-v_-)=\frac{1}{Q}=
\tfrac{1}{2}
\frac{\rd}{\rd Z}\left(
\log \left(\frac{\hat{\tau}_a(Z)^2}{\tau_a(Z)}\right) - 
\log \left(\frac{\hat{\tau}_{-a-3}(Z)^2}{\tau_{-a-3}(Z)}\right)
\right), 
$$  
and then using (\ref{tauids}) together with (\ref{Zscahod}) this gives 
$$ 
\rd z = Q^{-1}\, \rd Z = \tfrac{1}{2} 
\rd\, \log \left(\frac{\tau_{a+3}(Z)}{\tau_a(Z)}\right) 
,
$$ 
whence (\ref{zvntaus}) follows by integrating, with $a$ fixed in terms of $\al$ by (\ref{alfix}). 
\end{prf} 

\begin{exa}\label{vnsimexa}
As already mentioned, the ODE  (\ref{dp3rdred}) has a sequence of particular solutions that are algebraic in $Z$, at the 
parameter values $a=3n$ with $n\in\Z$; the first few are presented in Table 2 of \cite{bh}. 
For illustration of the preceding theorem, we consider the simplest of these, which is 
given by $P=P_0(Z)$ with 
$$
P_0=(Z/3)^{1/4}, \qquad a=0.  
$$
Putting this into (\ref{Ra}) and (\ref{alfix}) produces a corresponding solution $R=R_0(Z)$ of (\ref{RPV}), where 
$$ 
R_0=\left(\tfrac{Z}{3}\right)^{-1/4}\left( \Big(\tfrac{Z}{3}\Big)^{3/4}-\tfrac{1}{4} \right), \qquad \al=\tfrac{1}{2}, 
$$ 
which in turn leads to $Q=Q_0(Z)$ obtained from  (\ref{QRal}) as 
$$ 
Q_0=\left(\tfrac{Z}{3}\right)^{-1/2}\left( \Big(\tfrac{Z}{3}\Big)^{3/4}-\tfrac{1}{8} \right). 
$$ 
Upon applying the formula (\ref{UQRPV}), we take the plus sign, so that $U=R_0/\sqrt{Q_0}$; 
and, rather than computing the tau functions in (\ref{zvntaus}), we can directly calculate $z(Z)$ as the 
integral $z=\int Q_0(Z)^{-1}\, \rd Z+\,$const.  The resulting parametric solution of (\ref{vnsimred}) 
is more conveniently expressed by replacing $Z$ with the parameter 
$
 \eta = 4(Z/3)^{\sfrac{3}{4}}
$,  
corresponding to the independent variable for 
Painlev\'e III, as in 
(\ref{hatwP}),  
so that (up to an arbitrary choice of constant in $z$) it takes the form 
\beq\label{Uzparamvn}
U=\tfrac{1}{\sqrt{2}}\left(\frac{\eta-1}{\sqrt{2\eta -1}}\right), \qquad z=\eta +\tfrac{1}{2}\log(2\eta-1). 
\eeq 
To check that this agrees with the formula for $z$ in the above theorem, we can use the first two entries in Table 2 of \cite{bh} 
(replacing $\ze\to\eta$ therein), to read off the first two algebraic solutions of (\ref{dp3rdred}) in terms of $\eta$ as 
$$ 
P_0=(\eta/4)^{1/3}, \qquad P_3 = (\eta/4)^{1/3}\left(\frac{2\eta-3}{2\eta-1}\right). 
$$
Then substituting the above into (\ref{Vkka}) for $a=0,3$, and rewriting everything in terms of $\eta$ instead of $Z$, 
the two reduced Kaup-Kupershmidt fields are found as 
$$ 
V_0= 2^{\sfrac{1}{3}}\eta^{-\sfrac{8}{3}}\Big(\tfrac{7}{18}-\tfrac{1}{2}\eta^{2}\Big), \qquad 
V_3 = -2^{\sfrac{1}{3}}\eta^{-\sfrac{8}{3}}\frac{\Big(2\eta^4+2\eta^3+\tfrac{53}{18}\eta^{2}+\tfrac{14}{9}\eta-\tfrac{7}{18}\Big)}{(2\eta-1)^2}, 
$$ 
and then integrating twice with respect to $Z$ and using (\ref{taukksk}), 
the corresponding tau functions are also written conveniently in terms of the 
same independent variable for 
Painlev\'e III, up to a choice of gauge, as 
$$ 
\tau_0 = \eta^{-\sfrac{7}{36}}\exp\Big(-\tfrac{1}{4}\eta^2\Big), 
\qquad 
\tau_3 = \eta^{-\sfrac{7}{36}}(2\eta -1)\exp\Big(-\tfrac{1}{4}\eta^2+2\eta\Big),
$$  
so that calculating $\sfrac{1}{2}(\log\tau_3-\log\tau_0)$ from (\ref{zvntaus}) indeed reproduces the expression for 
$z$ in (\ref{Uzparamvn}). 

Taking real $\eta>\sfrac{1}{2}$ ensures that  $U(z)$ is real-valued. 
A plot of this solution appears in Fig.6. 
The behaviour as $\eta$ approaches $\sfrac{1}{2}$ from above is 
$$ 
\eta \rightarrow \tfrac{1}{2}+ \implies z\to -\infty,\quad U\to -\infty, 
$$ 
with leading order asymptotics described by  
\beq\label{vnexpasy}
U\sim- \frac{1}{2\sqrt{2}}\, \frac{1}{\sqrt{2\eta-1}}, \quad 
z\sim \tfrac{1}{2}\log(2\eta -1)+\tfrac{1}{2} \implies 
U\sim - \frac{1}{2\sqrt{2}}\,e^{-\big(z-\sfrac{1}{2}\big)}. 
\eeq
For large $\eta$ the behaviour is 
$$ 
\eta \rightarrow \infty \implies z\to \infty,\quad U\to \infty, 
$$ 
with leading order asymptotics
$$ 
U\sim \tfrac{1}{2}\sqrt{\eta}, \quad z\sim \eta \implies U\sim\tfrac{1}{2}\sqrt{z}. 
$$ 
However, the latter does not provide a particularly accurate approximation to the solution. 
Much greater accuracy can be achieved by reverting the equation for $z$ in  (\ref{Uzparamvn}) as 
$\eta = z-\sfrac{1}{2}\log(2\eta-1)$, using this to generate an expansion 
$$ 
\eta = z-\tfrac{1}{2}\log z-\tfrac{1}{2}\log 2+o(1) 
$$ 
where the omitted terms above are a double series in powers of $\log(z)$ and $z^{-1}$, 
and substituting into the formula for $U$ in terms of $\eta$ then gives 
\beq\label{Ubetter} 
U\sim \tfrac{1}{2}\sqrt{z}\Big(
1-\tfrac{1}{4}z^{-1}\log z-\tfrac{1}{4}(3+\log 2)z^{-1}
\Big), \qquad z\to\infty, 
\eeq 
omitting terms inside the big brackets above that are $o(z^{-1})$. In Fig.7 we have overlaid plots of the asymptotic approximations 
(\ref{vnexpasy} (blue) and (\ref{Ubetter}) (red) on top of part of the plot from Fig.6, which show 
quite good agreement 
even for relatively modest magnitudes of $z$ when it is negative/positive, respectively. 
\end{exa}

\begin{figure}
 \centering
\label{Uzvnfigas}
\epsfig{file=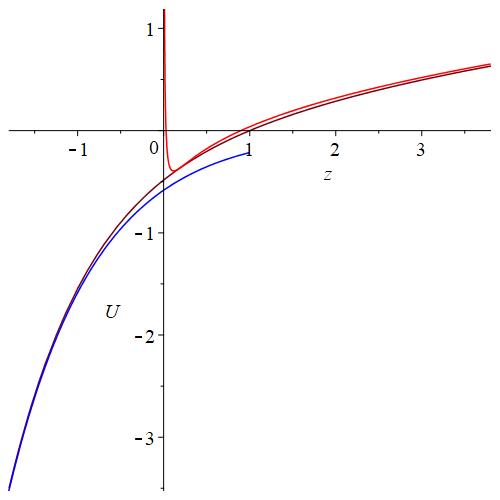, height=2.7in, width=2.7in}
\caption{Plot of $U(z)$ 
compared with asymptotic formulae for the parametric solution 
(\ref{Uzparamvn}).
}
\end{figure} 

In most of our analysis we have made the implicit assumption that $\al\neq 0$, which was used in the derivation 
of the ODE (\ref{RPV}) for $R$.   The case $\al=0$ (separable solutions of Novikov's equation) corresponds to  integrating (\ref{QRdZ}) 
with the integration constant in  (\ref{QRal})
being zero. This implies that $R^2-Z/3=0$, which can be regarded as a singular solution 
of the ODE with $\al=0$, because both the denominator and the numerator inside the large brackets on the right-hand side 
of (\ref{RPV}) vanish.
Then  $Q(Z)$ satisfies the second order ODE 
\beq\label{QODE} 
\frac{(Q^{1/2})_{ZZ}}{Q^{1/2}}+\frac{1}{Q^2}=-\frac{1}{4Z^2}\pm \frac{1}{\sqrt{\tfrac{Z}{3}}}, 
\eeq 
obtained  from substituting $R=\pm\sqrt{\sfrac{Z}{3}}$ into the right-hand side of (\ref{VhatQR}). 
The above equation reduces to a Riccati equation for $v$ defined by (\ref{vsym}), taking $v=v_+$ without 
loss of generality, namely 
\beq\label{ricc}
\frac{\rd v}{\rd Z}-v^2 =\frac{1}{4Z^2}\mp \frac{1}{\sqrt{\tfrac{Z}{3}}},
\eeq 
and given the 
solution $v(Z)$ of the latter, $Q$ is then found from the solution of the inhomogeneous linear equation 
\beq\label{Q1st} 
\tfrac{1}{2}\frac{\rd Q}{\rd Z}+v Q= 1.
\eeq 
The other solutions of the ODE  with $\al=0$ are not directly relevant to the scaling similarity solutions of 
Novikov's equation, but they have an indirect relevance  via the connection to the solutions at parameter values  
$\al=3n$ for non-zero integers $n$, corresponding to the solutions of (\ref{dp3rdred}) 
for $a=3\big(n+\sfrac{1}{2}\big)$ that are related to one another by the B\"acklund transformation (\ref{shifta}).\footnote{As pointed out in \cite{bh}, for all such values of $a$ there is a one-parameter family of 
special solutions, given in terms of Bessel functions, corresponding to classical solutions 
of Painlev\'e III.}

In order to describe the solutions with $\al=0$ more explicitly, we set 
$$
v=-\frac{\rd}{\rd Z}\log \psi, 
$$
in the Riccati equation (\ref{ricc}), 
which transforms it to the Schr\"odinger equation 
\beq\label{psiSchro}
\frac{\rd^2\psi}{\rd Z^2}+\left( \frac{1}{4Z^2}\mp \frac{1}{\sqrt{\tfrac{Z}{3}}}\right)\psi=0
 .
\eeq
Then, upon changing variables according to  
$$ 
\psi(Z)=Z^{\sfrac{1}{2}}\phi(\eta), \qquad 
 \eta = 4(Z/3)^{\sfrac{3}{4}}, 
$$
once again using the independent variable $\eta$ for 
Painlev\'e III, the equation (\ref{psiSchro}) becomes 
$$
\eta^2 \frac{\rd^2\phi}{\rd \eta^2}+\eta \frac{\rd\phi}{\rd \eta}\mp \eta^2\phi=0, 
$$ 
which is solved in terms of   Bessel/modified Bessel functions of order 0, depending on the sign. In particular, 
with the plus sign above, which corresponds to the case $R=-\sqrt{Z/3}$, this implies that the general solution of (\ref{psiSchro}) 
can be written as 
\beq\label{psisol} 
\psi(Z) = A Z^{1/2} J_0\Big(4(\tfrac{Z}{3})^{\sfrac{3}{4}}\Big)+BZ^{1/2} Y_0\Big(4(\tfrac{Z}{3})^{\sfrac{3}{4}}\Big), 
\eeq 
for arbitrary constants $A,B$. 
By replacing $v$ in (\ref{Q1st}) in terms of $\psi$, this reduces to a quadrature for $Q$, namely 
\beq\label{Qquad} 
Q(Z) = 2\psi(Z)^2 \left(\int^Z \frac{\rd s}{\psi(s)^2}+C\right), 
\eeq 
for another arbitrary constant $C$. (Observe that the formula (\ref{Qquad}) only depends on the ratio $A/B$, 
so overall this gives two arbitrary constants in the general solution of (\ref{QODE}), as required.) 

For completeness,  the case $\al=0$ is summarized as follows.

\begin{thm}\label{alzero} 
The solutions of the ODE (\ref{vnU3rd}) with $\al=0$, which for $z=x$ 
correspond to the separable solutions 
(\ref{sep}) of Novikov's equation (\ref{novikov}),  
are given parametrically in the form  $U=U(Z)$, $z=z(Z)$,  
with  
\beq\label{Ualzero} 
U(Z)=\pm\sqrt{\frac{Z}{3Q(Z)}}
, 
\qquad 
z(Z)=\int^Z \frac{\rd s}{Q(s)}+\mathrm{const},
\eeq 
where $Q(Z)$ is a solution of the ODE (\ref{QODE}), given by the 
quadrature (\ref{Qquad}) with $\psi$ specified as in (\ref{psisol}), or by an 
analogous formula with modified Bessel functions of order 0. 
\end{thm} 


\section{Conclusions} 

\setcounter{equation}{0}

In this paper we have found parametric formulae for the scaling similarity solutions 
of two integrable peakon equations with cubic nonlinearity, namely (\ref{mch}) and (\ref{novikov}). 
In both cases, by applying the similarity reduction to suitable reciprocal transformations, and 
using Miura maps between negative flows of appropriate integrable hierarchies,   
we have shown that these parametric solutions are related to 
Painlev\'e III transcendents, for specific values  of the parameters 
$\tilde{\al}, \tilde{\be}, \tilde{\gam}, \tilde{\delta}$ in (\ref{piii}). 
More precisely, the  scaling similarity solutions  of the mCH equation 
(\ref{mch}) are related to the same case of Painlev\'e III that arises from 
the Camassa-Holm equation (\ref{ch}), while for Novikov's equation  (\ref{novikov})
such solutions are related to the case of Painlev\'e III that is 
associated with an analogous reduction of 
 the Degasperis-Procesi equation (\ref{dp}).

The scaling similarity solutions of the mCH equation (\ref{mch}) 
have been written parametrically in terms of solutions of the ODE (\ref{F2nddeg}), which is of second 
order and second degree. 
The systematic study of such equations was initiated in \cite{cs}, although to the best of our knowledge 
there is still no complete classification of second order, second degree equations with the 
Painlev\'e property.  Certain particular equations of this type, the so-called sigma forms 
of the Painlev\'e equations, which  are the equations satisfied by Okamoto's Hamiltonians \cite{o}, play an 
important role 
in both theory and applications. However,  the ODE (\ref{F2nddeg}) is of a different kind, 
since the Hamiltonians are  quadratic functions of the first derivatives of the solution 
of the corresponding Painlev\'e equation, whereas the transformation (\ref{FPdef}) is linear 
in $\sfrac{\rd P}{\rd Z}$, with $P(Z)$ being a solution of Painlev\'e III. 

For the case of Novikov's equation (\ref{novikov}), the scaling similarity solutions are 
expressed parametrically in terms of solutions of the ODE (\ref{RPV}), 
which is equivalent to the Painlev\'e V equation with a particular choice of parameters, 
and arises via  reduction of the negative Sawada-Kotera flow (\ref{skneg}).  
The equation (\ref{RPV}) has a one-to-one correspondence with the ODE (\ref{dp3rdred}) 
obtained via the scaling similarity reduction for the negative Kaup-Kupershmidt flow  
(\ref{VTdp}), which in turn is equivalent to another particular case of Painlev\'e III. 
The correspondences and B\"acklund transformations between the solutions of  
(\ref{dp3rdred}) and (\ref{RPV}) have been constructed using 
properties of these solutions that are naturally inherited from the two Miura maps in 
(\ref{kkskmiura}), which relate the  Kaup-Kupershmidt and Sawada-Kotera PDE hierarchies 
to the same underlying modified hierarchy. However, implicit in our construction is the fact that there is  
a negative flow in the latter hierarchy, which should be given by a PDE of third order for the modified field $v=v(X,T)$, 
while at the level of the scaling similarity reductions there must be an ODE of second order for the reduced 
variable $v(Z)$. It has not been  necessary to write them down here, but computer algebra calculations show 
that these equations are somewhat unwieldy: 
the modified PDE for $v(X,T)$ is of second degree in the highest derivative that appears, namely $v_{XXT}$, 
while the reduced ODE for $v(Z)$ is of  third degree in its second derivative $v_{ZZ}$, so we have thought it best to 
leave a more detailed discussion of these matters for elsewhere.

\noindent \textbf{Acknowledgments:} LEB was supported by a PhD studentship from SMSAS, Kent. The research of ANWH was funded in 2014-2021 by 
Fellowship EP/M004333/1 from the
Engineering \& Physical Sciences Research Council, UK, with a UKRI COVID-19 Grant Extension, and is currently supported by 
grant 
 IEC\textbackslash R3\textbackslash 193024 from the Royal Society. 
The  inception of this research project was made possible 
thanks to support from the Royal Society and the Department of Science \& Technology (India) 
going back to 2005, which allowed Senthilvelan to visit Hone under the India-UK Science Network and 
Scientific Seminar schemes.  
Conflict of Interest: The authors declare that they have no
conflicts of interest.


\end{document}